\documentclass{aa}  

\usepackage{subcaption}
\usepackage{graphicx}

\usepackage{txfonts}
\usepackage{physunits}
\usepackage{xcolor}
\usepackage{placeins}

\usepackage{siunitx}
\usepackage{booktabs}

\ifdefined\unit\else
  \ifdefined\NewCommandCopy
    \NewCommandCopy\unit\si
  \else
    \NewDocumentCommand\unit{O{}m}{\si[#1]{#2}}
  \fi
\fi

\usepackage{natbib}
\usepackage[breaklinks=true,hidelinks]{hyperref}

\bibpunct{(}{)}{;}{a}{}{,}
\begin{document}

   \title{A Three-Dimensional Tomographic Reconstruction of the Galactic Cosmic-Ray Proton Density}

    \author{
      Hanieh Zandinejad\inst{1,2}
      \and
      Jakob~Roth\inst{3}
      \and
      Vo~Hong~Minh~Phan\inst{7} 
      \and
      Gordian~Edenhofer\inst{1}
      \and
      Philipp~Frank\inst{4}
      \and
      Philipp~Mertsch\inst{5}
      \and
      Ralf~Kissmann\inst{6}
      \and   
      Andrés~Ramírez\inst{6}
      \and
      Laurin~Söding\inst{1,5}
      \and
      Torsten~A.~Enßlin\inst{1,2}
      }

    \institute{
      Max Planck Institute for Astrophysics,
      Karl-Schwarzschild-Straße 1, 85748 Garching bei München, Germany \\\email{haniehzd@mpa-garching.mpg.de}
      \and
      Ludwig Maximilian University of Munich,
      Geschwister-Scholl-Platz 1, 80539 München, Germany
      \and
      Max Planck Computing and Data Facility,
      Gießenbachstraße 2, 85748 Garching, Germany
      \and
      Kavli Institute for Particle Astrophysics and Cosmology,
        Stanford University,
        Stanford, CA 94305, USA
      \and
      Institute for Theoretical Particle Physics and Cosmology (TTK),
      RWTH Aachen University, 52074 Aachen, Germany 
      \and
      Institut für Astro- und Teilchenphysik, Universität Innsbruck,
      Technikerstr. 25, 6020 Innsbruck, Austria
      \and
      Sorbonne Université, Observatoire de Paris, PSL Research University,
      LERMA, CNRS UMR 8112, 75005 Paris, France
    }


 
  \abstract
{Cosmic rays (CRs) are a ubiquitous non-thermal component of the interstellar medium (ISM). A data-driven three-dimensional (3D) map of their distribution is essential for understanding CR transport and constraining the spatial distribution of their sources.}
{We aim to reconstruct the 3D spatial distribution of the Galactic cosmic-ray proton (CRp) density.}
{We model the diffuse gamma-ray emission arising from inelastic hadronic interactions between CRps and interstellar gas. Using a map of dust-correlated diffuse gamma-ray emission based on ten years of \textit{Fermi}-LAT data together with a three-dimensional gas density model, we infer the spatial CRp distribution via a morphological matching approach. The logarithmic CRp density field is described by a Gaussian process on a spherical times radial grid, and both the field and its correlation structure are inferred simultaneously using Iterative Charted Refinement. The posterior distribution of the three-dimensional CRp density field is approximated using geometric variational inference.}
{The reconstructed CRp density exhibits a smooth but spatially structured distribution with a limited dynamical range across the Galactic disk. We find a moderate enhancement of the CRp density toward the inner Galaxy. The inferred normalization at the Solar position is consistent with local CR measurements by the AMS--02 instrument.}
{}


   \maketitle
%
\section{Introduction}

Cosmic rays (CRs) are a key non-thermal component of the interstellar medium (ISM) and play a fundamental role in Galactic astrophysics. They contribute significantly to the total pressure balance of the ISM, influence the ionization state and chemistry of dense clouds, and drive turbulence and large-scale outflows that regulate the evolution of the Milky Way \citep{Grenier15, Zweibel13}. Despite their importance, the global spatial distribution of CRs in three dimensions remains poorly constrained. While local measurements near Earth provide detailed information on the CR spectrum and composition, they represent only a single point in the Galaxy. Understanding how the CR density varies across the Galactic disk and halo is crucial for quantifying their feedback on the ISM, identifying regions of active acceleration, and testing models of CR transport and confinement. 

The interaction of CRs with the components of the ISM gives rise to diffuse gamma-ray emission, which serves as a powerful indirect tracer of CR populations. Three main processes contribute to the Galactic diffuse gamma-ray emission: (i) hadronic interactions between CR hadrons and ambient gas nuclei, leading to the production and decay of neutral pions ($\pi^{0} \rightarrow 2\gamma$); (ii) inverse Compton scattering of CR electrons and positrons off the interstellar radiation field; and (iii) bremsstrahlung from CR electrons interacting with the interstellar gas \citep{Aharonian00, Ruszkowski23}. In the energy range above $\sim 1$ GeV and up to several tens of GeV, the hadronic component largely dominates the total gamma-ray flux, making it a probe of the CR proton (CRp) density and spectrum across the Galaxy.

The pion-decay gamma-ray flux can be modeled by convolving the distribution of CRps with the gas density along the line of sight. Therefore, measurements of the diffuse Galactic gamma-ray emission obtained by instruments such as the Fermi Large Area Telescope (LAT) provide essential constraints on the large-scale distribution of CRs and their propagation properties. The observed gamma-ray intensity profile across the Galactic disk reveals variations in both the CR density and spectral index, suggesting spatially dependent transport properties or localized CR sources (e.g. \citealp{Yang16, Pothast2018, Recchia16}).

The Galactic CR distribution has been constrained through a combination of modeling and observational approaches. Numerical propagation models such as GALPROP, DRAGON, and PICARD \citep{Strong07, Evoli20, Kissmann14} simulate CR transport by solving the diffusion--loss equation under assumed source distributions, diffusion coefficients, and boundary conditions. These models successfully reproduce local CR spectra and large-scale gamma-ray emission, but rely on simplifying assumptions that limit their predictive power for three-dimensional spatial variations. Moreover, persistent residuals between modeled and observed diffuse gamma-ray emission highlight limitations in current descriptions of CR transport and target distributions, motivating more flexible, data-driven approaches \citep{Ackermann2012diffuse, buschmann2020foreground, karwin2023improved}.

Complementary observational methods use gamma-ray data to infer the CR distribution empirically. One class of studies employs ring-model analyses, in which the Galaxy is divided into concentric Galactocentric rings and the gamma-ray emissivity is fitted in each region. This technique allows the characterization of the CR density gradient across the Galactic disk and has revealed a progressive hardening of the CR spectrum toward the inner Galaxy \citep{Pothast2018, Cerri17, Vecchiotti21}. Such findings have stimulated extensive theoretical work exploring the impact of anisotropic diffusion and spatially dependent propagation coefficients \citep{Gabici19, Evoli20}. 

A second approach focuses on molecular clouds (MC), which provide localized measurements of the CR density in specific regions of the Galaxy. Massive molecular clouds act as passive barometers of the ambient CR population: their dense gas content serves as an effective target for CR--gas interactions, and the resulting gamma-ray flux directly reflects the CR density at the cloud location \citep{Yang13, Neronov11, Aharonian2020, Peron21}. This method has been successfully used to probe the spatial variations and degree of uniformity of the Galactic CR ``sea'' in the local interstellar medium and in distant Galactic regions, including the inner Galaxy \citep{Prokhorov18, Peron21, Albert21}.

These studies generally find that the CR density in the outer and local Galaxy is consistent with the local interstellar spectrum measured near Earth, while enhanced CR densities are observed toward the inner Galaxy and the Galactic center. These deviations may indicate the presence of additional CR sources, enhanced CR confinement, or uncertainties in the underlying gas mass distribution used in the analyses. For example, \citet{Peron21} analyzed nine molecular clouds located at Galactocentric distances of approximately 1.5--4.5~kpc and found that, while many clouds exhibit CRp densities comparable to the local interstellar value, the emissivity distribution inferred from diffuse gamma-ray analyses shows a pronounced enhancement around a Galactocentric radius of $\sim$4~kpc. The CR energy density in this region can exceed the local value by roughly 20--50\%, with some individual molecular clouds displaying even larger deviations. These results suggest that the CR population in the inner few kiloparsecs is not uniform and may be influenced by nearby accelerators, spatial variations in the propagation conditions, or uncertainties in the gas mass distribution, leading to an increased confinement of CRs in the central molecular zone and the inner Galactic disk.

Recent gamma-ray observations from \textit{Fermi}-LAT and ground-based instruments such as H.E.S.S. and HAWC have further advanced our understanding of the Galactic CR distribution \citep{Sinha21, Roy24, Vecchiotti25}. However, the interpretation of the observed gradients remains debated. It is still unclear whether the variations arise from a higher concentration of CR accelerators in the inner Galaxy or from spatial changes in the CR propagation regime. Disentangling these effects requires a combination of large-scale emissivity modeling and localized CR measurements based on molecular clouds.

In this work, we present a fully data-driven, three-dimensional tomographic reconstruction of the Galactic CRp density based on diffuse gamma-ray observations, with minimal reliance on assumed global propagation models. Our approach combines diffuse gamma-ray emission with a three-dimensional gas distribution \citep{gas_25} within a tomographic framework to constrain spatial variations of the CRp density across the Galaxy.

We model the CRp density as a three-dimensional non-parametric field using a Gaussian process prior and infer it within a Bayesian framework. This allows us to reconstruct the CRp distribution directly from the gamma-ray data, while simultaneously quantifying uncertainties and avoiding the need for predefined radial or axisymmetric models. By explicitly modeling the hadronic gamma-ray emission induced by CR--gas interactions and combining it with an adopted three-dimensional gas model, we infer the spatial variations of the CRp density that are consistent with the gamma-ray data under this gas assumption.

This paper is structured as follows. In Section~2, we describe the methodology adopted in this work. We introduce the construction of the diffuse gamma-ray emission maps and the three-dimensional gas distribution used in the tomographic forward-modeling framework. We then present the modeling of gamma-ray emission arising from hadronic CRp interactions, including the underlying CRp distribution and pion-decay gamma-ray production. The Bayesian framework employed to infer the CRp density, together with the adopted prior model and inference algorithm, is detailed next, followed by a validation of the method using synthetic data. The results of applying this framework to gamma-ray observations are presented in Section~3. In Section~4, we discuss the implications of our findings for the Galactic CR distribution and transport properties, and compare them with previous studies. Finally, Section~5 summarizes our main conclusions and outlines perspectives for future work.
\begin{figure*}[tbp]
    \centering
    \includegraphics[width=0.7\linewidth]{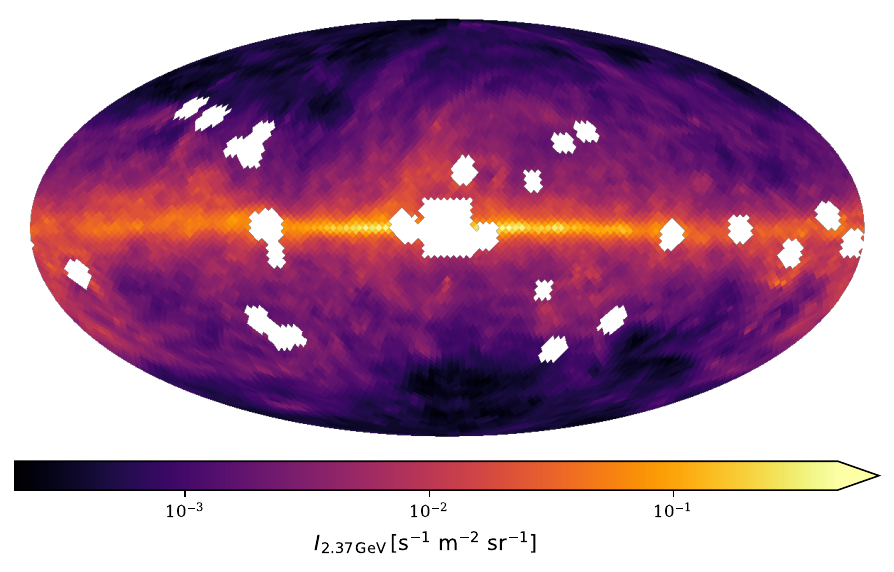}
    \caption{
        Masked dust-correlated diffuse gamma-ray emission map in the energy range between
        1.78 and 3.16~GeV used as input data for the inference. The map --as well as any other all sky map in this publication-- is shown in a Mollweide projection and downgraded to
        an angular resolution corresponding to $N_{\text{side}} = 32$.}
    \label{fig:gamma_data_masked}
\end{figure*}
\section{Method}
To construct the 3D distribution of the CR density, we develop a statistical forward model for the CR density and the resulting gamma-ray emission, enabling the inference of the CR density map by means of Bayesian inference. This section is structured as follows: in Section~\ref{data}, we introduce the diffuse gamma-ray sky map data. In Section~\ref{gas map}, we present the 3D gas density map used as input to the tomographic forward-modeling framework. In Section~\ref{gamma-ray source model}, we discuss the model describing the gamma-ray emission arising from hadronic interactions between CRps and nucleons of the ISM. We then provide an overview of the Bayesian inference framework and describe the prior model for the CR density field together with the discretization scheme and imposed correlation structure in Section~\ref{prior model}. In Section~\ref{likelihood}, we introduce the likelihood and noise model used to relate the predicted gamma-ray emission to the observed diffuse gamma-ray data. In Section~\ref{inference}, we describe the forward model connecting the CR density field to the observed gamma-ray data and summarize the adopted inference algorithm. Finally, in Section~\ref{validation}, we validate the full inference pipeline on synthetic data to demonstrate its performance and assess its reconstruction capabilities under controlled conditions.
\begin{table}[t]
\centering
\small
\caption{Locations of masked regions in Galactic coordinates. Circular masks are applied to the listed objects. The Galactic center is masked as a rectangular region.}
\label{tab:masked_regions}
\begin{tabular}{lcc}
\toprule
Category & \multicolumn{2}{c}{$(l,b)\,[^\circ]$} \\
\midrule

Magellanic Clouds 
& (280.47, $-32.89$) & (302.80, $-44.33$) \\

\midrule

Dust peaks (SP23)
& (262.97, $-2.39$) & (88.59, 24.62) \\
& (195.47, 4.78) & (85.78, $-38.68$) \\
& (77.34, $-38.68$) & (185.63, $-5.98$) \\
& (322.03, 16.96) & (292.50, 34.23) \\
& (98.44, 28.63) & (305.16, 30.00) \\
& (74.53, $-9.59$) & (135.00, 43.41) \\
& (316.41, $-22.02$) & (90.00, 34.23) \\

\midrule

Pulsars
& (90.0, 26.0) & (162.7, $-16.0$) \\
& (78.2, 2.1) & (75.2, 0.1) \\
& (80.2, 1.0) & (76.1, $-0.2$) \\
& (352.6, 20.2) & (343.1, $-2.7$) \\
& (7.0, $-1.2$) & (18.6, $-0.4$) \\
& (6.4, 4.9) & (350.6, $-4.7$) \\
& (4.9, $-1.2$) & (17.2, $-0.3$) \\
& (20.1, 1.3) & (263.6, $-2.8$) \\
& (195.1, 4.3) & (184.6, $-5.8$) \\
& (233.5, $-0.3$) & (210.4, $-9.3$) \\
& (90.0, $-33.0$) & (85.0, $-40.0$) \\
& (120.0, 40.0) & \\

\midrule

Galactic center 
& \multicolumn{2}{l}{Rectangular mask: $|l|\leq 10^\circ,\ |b|\leq 10^\circ$} \\

\bottomrule
\end{tabular}
\end{table}
\subsection{Diffuse Gamma-ray Emission map} \label{data}

Fermi Large Area Telescope (LAT) measurements provide the currently most extensive dataset of gamma-ray sky observations~\citep{2009ApJ...697.1071A}. A rich set of analysis methods has been developed to study these data, most of which rely on predefined emission templates to model foreground components, as demonstrated in numerous analyses (e.g.~\citealp{Ackermann2012diffuse, Acero2016, buschmann2020foreground, karwin2023improved}). More recent developments based on information field theory (IFT) enable a simultaneous reconstruction and decomposition of the gamma-ray sky into multiple emission components in a largely template-free manner \citep[][hereafter referred to as Scheel-Platz23]{2023A&A...680A...2S}. In this approach, emission components are modeled based on their morphology and inferred by exploiting their spatial and spectral correlation structures.

The reconstruction by Scheel-Platz23 separates the gamma-ray sky into diffuse emission and point-source contributions. For each of these components, the inferred energy-integrated gamma-ray photon flux density $I_{ij}$, with units $(\unit{\second \square\metre \steradian})^{-1}$, is provided and defined as
\begin{equation}\label{gamma flux dens}
    I_{ij} = \frac{1}{|\Omega_j|} \int_{E_i}^{E_{i+1}} dE \int_{\Omega_j} dx \ \Phi(x,E),
\end{equation}
where $\Phi(x,E)$ denotes the time-averaged photon flux density as a function of photon origin direction $x$ and energy $E$, integrated over the energy range 0.56--316~GeV and the first ten years of LAT operation. Here, $\Omega_j$ is the solid angle of the $j$-th sky pixel, and the energy bin boundaries are chosen equidistantly on a logarithmic energy scale. The sky is discretized using the HEALPix scheme with $N_{\text{side}} = 128$, and the energy dimension is sampled with four bins per decade.

In addition to separating point sources from diffuse emission, Scheel-Platz23 further decomposes the diffuse component using a template-informed model based on the \textit{Planck} 545~GHz thermal dust emission map. This template is supplemented by a free-form field that allows for deviations from the dust morphology. As a result, the diffuse gamma-ray emission is separated into a dust-correlated component and a dust-uncorrelated component, the latter capturing emission processes whose target distributions are not traced by the dust template, such as inverse Compton scattering of cosmic-ray electrons on the interstellar radiation field or contributions from large-scale structures like the Fermi bubbles \citep{Su2010}.

In this work, we use the dust-correlated diffuse gamma-ray emission component in the energy range between 1.78 and 3.16~GeV as our primary dataset. The map is downgraded to a lower angular resolution corresponding to $N_{\text{side}} = 32$, consistent with the resolution adopted for the three-dimensional gas distribution and the CRp density reconstruction.

To minimize contamination from emission components that are not adequately described by our hadronic diffuse emission model, we apply a spatial mask to the gamma-ray data. In particular, regions associated with prominent extragalactic gamma-ray sources, namely the Large and Small Magellanic Clouds (LMC and SMC), are excluded, as their emission originates outside the Milky Way and is incompatible with a Galactic CR density reconstruction. In addition, localized regions around bright point sources that are not perfectly separated from the diffuse emission in Scheel-Platz23 reconstruction are masked. This is especially relevant for very bright sources with complex or extended emission morphologies, such as the Vela region, where residual point-source contamination can bias the inferred diffuse gamma-ray intensity.

In addition to the masking of the LMC, SMC, and localized bright point-source regions, we apply a further data-driven masking procedure guided by the properties of the dust-correlated modification field presented by Scheel-Platz23. Their Fig.~16 highlights sky locations where a significant multiplicative correction to the thermal dust template is required in order to account for residual emission. These regions indicate locations where the diffuse emission model departs strongly from the dust-correlated morphology and are frequently associated with bright or complex gamma-ray sources. Several of the additionally masked regions coincide with prominent nearby gamma-ray sources and complex emission structures, including the Vela region, the Cygnus region, and areas near the Galactic center. The adopted masking regions and their corresponding angular scales are summarized in Table~\ref{tab:masked_regions}. Circular masks with radii of $5^\circ$ are applied to the LMC, SMC, and selected bright source regions, while the dust-template modification peaks are masked using circular regions of radius $4^\circ$. In addition, a rectangular region around the Galactic center with extent $|l| \leq 10^\circ$ and $|b| \leq 10^\circ$ is excluded in the main analysis.

To reduce the impact of residual non-hadronic emission on our reconstruction, we identify regions of pronounced template modification and mask the corresponding sky pixels. We additionally cross-reference these locations with the Fermi pulsar sky map to ensure consistency with known bright pulsars. Pixels where strong template modification spatially coincides with identified pulsars are excluded from the analysis, as such emission is not described by our hadronic CRp model.

Finally, we exclude a rectangular region around the Galactic center. In this direction, the underlying 3D gas distribution used in the tomographic forward-modeling framework is comparatively uncertain, which significantly distorts the reconstructed CRp density when this region is included. In particular, the resulting CRp map exhibits artificial structures that are not consistent with the large-scale morphology inferred elsewhere. To avoid biasing the reconstruction through these poorly constrained gas features, we therefore mask the Galactic-center region in the main analysis. The corresponding reconstruction without this mask is shown in Appendix~\ref{appendix_gc_mask} for reference.

The resulting masked gamma-ray map constitutes the dataset used throughout this work. Figure~\ref{fig:gamma_data_masked} shows the corresponding masked gamma-ray map.
\begin{figure*}[tbp]
    \centering
    \includegraphics[width=0.8\linewidth]{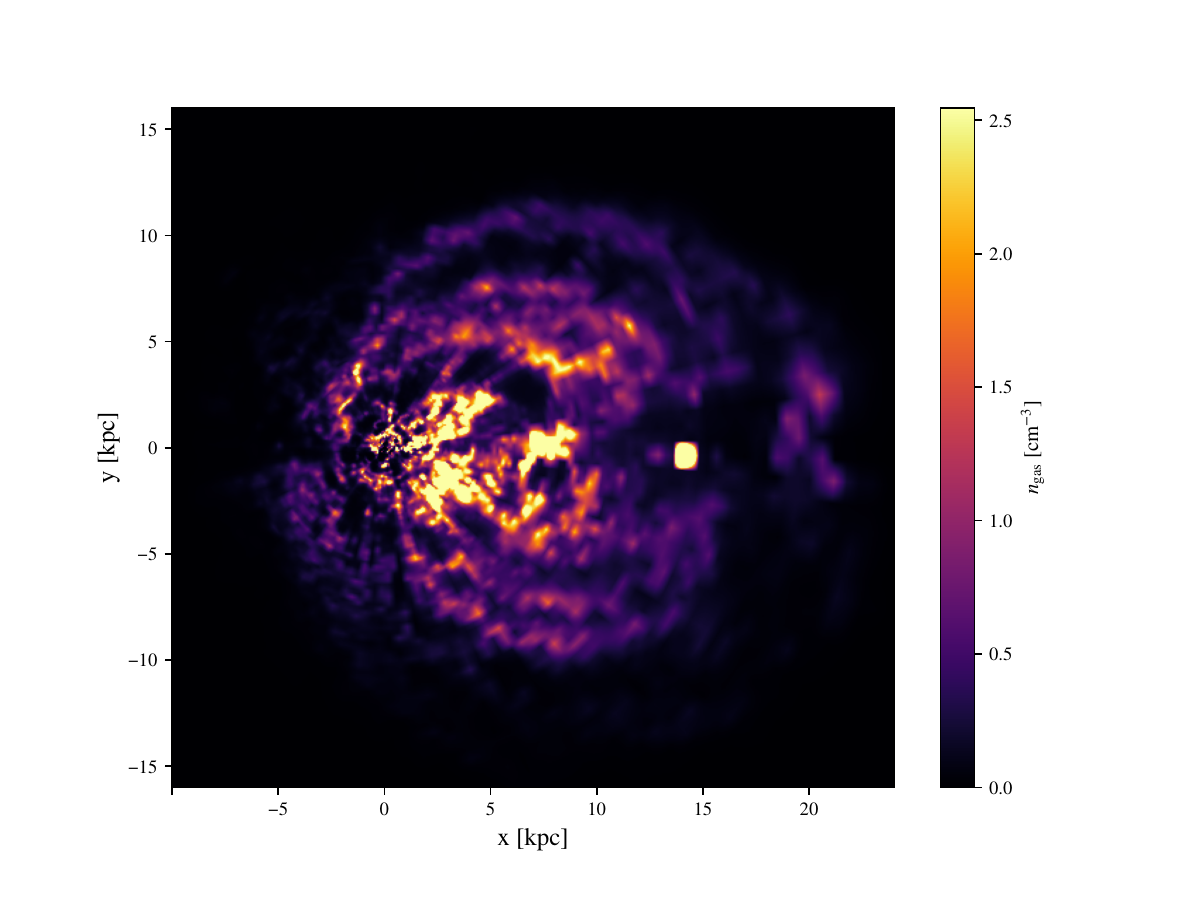}
    \caption{Top-down view of the total Hydrogen density $n_\text{gas}$ at $z=0$ inferred by Söding25. The position of the  Sun is at (0,0). Positive x points towards the Galactic centre. The corresponding resolution of the map depicted here differs from the original resolution of this map and is according to the resolution chosen for this work.}
    \label{gas_map}
\end{figure*} 
\subsection{Morphological Matching via a 3D Gas Map}\label{gas map}
While hadronic gamma-ray emission traces the distribution of CRps and their respective targets, the observed diffuse gamma-ray sky map does not provide direct distance information. Consequently, the reconstruction of the CRp density in three dimensions is only possible through the use of additional three-dimensional information about the target gas distribution. In this work, we refer to this tomographic forward-modeling procedure, which relates the observed diffuse gamma-ray emission to the three-dimensional gas distribution through a modeled CRp density field, as \textit{morphological matching}.

The recent developments in Galactic tomography resulting in 3D gas and dust maps of the Milky Way enables this \citep{gas_25,dust_edh,leike_dust,RWTH_gas_23,RWTH_gas_21,2022A&A...664A.174V,2019ApJ...887...93G}. In particular, we use the 3D gas map by \citet[][hereafter referred to as  Söding25]{gas_25}. The results provided by Söding25 consist of a joint reconstruction of the spatial distributions of HI and CO, their emission line-widths, and the Galactic velocity field using Bayesian inference. The resulting map consists of a set of posterior samples implicitly accounting for statistical uncertainties. In this work, we use the posterior mean of those samples. 

Although the primarily inferred quantities by Söding25 are the HI and CO distributions, the latter can be transformed to $\mathrm{H}_2$ via the commonly used linear relation between the $\mathrm{H}_2$ column density  $N_{H_2}$ and the velocity-integrated CO brightness temperature \citep{co_h_2_conversion}. Therefore, we can also define the total Hydrogen density as:
\begin{equation}
    n_{\text{H}} = n_{\text{HI}} + 2 \cdot n_{H_2}. 
\end{equation}

Hydrogen is the dominant baryonic component of the ISM and is present mainly in neutral atomic (HI) or molecular ($\mathrm{H}_2$) form. The ionized component, by contrast, is comparatively diffuse and spatially smooth, and is therefore expected to contribute less prominently to the dust-correlated gamma-ray emission component considered here.

For hadronic gamma-ray production, interactions involving heavier nuclei in both the CR population and the ISM also contribute to the total emission. In this work, however, we restrict the modeling to CR protons and hydrogen targets for simplicity. The additional contribution from heavier elements is accounted for through a nuclear enhancement factor of approximately $1.8$ \citep[see also the discussion below and][]{Kafexhiu:2014cua}. Furthermore, the hydrogen density inferred from the HI and $\mathrm{H}_2$ distributions is converted into a total nucleon density by applying a scaling factor that accounts for the helium abundance in the ISM. Figure~\ref{gas_map} shows a top-down view of the total hydrogen density in the $z=0$ slice from Söding25. The map is downgraded to the angular resolution adopted in this work, corresponding to $N_{\mathrm{side}} = 32$.
\subsection{Gamma-ray emission from Hadronic CRp Interactions}\label{gamma-ray source model}
In order to reconstruct the CRp density, we need a description of the gamma-ray emission resulting from CRp interactions with ISM nucleons. For this, we follow the terminology of \citet{2004A&A...426..777P} and define the omnidirectional differential source density $s(\textbf{r},E)$ as the number of gamma-ray photons produced per unit volume, time, and energy, and the integrated number density production rate $\lambda(\textbf{r})$ as the total gamma-ray production rate per unit volume in the energy range $[E_1, E_2]$:
\begin{equation}
    s(\textbf{r},E) = \frac{dN}{dV \, dt \, dE},
\end{equation}
\begin{equation}
    \lambda(\textbf{r}) = \int_{E_1}^{E_2} dE \, s(\textbf{r},E),
\end{equation}
where $N$ denotes the number of gamma-ray photons.

\subsubsection{Cosmic Ray Proton Distribution \& Hadronic Pion Production}\label{crp distribution}
The differential number density distribution of CRp as a function of momentum can often be described and approximated by a single power law \citep{ensslin2007}
\begin{equation}\label{cr momentum dist}
    f_{\text{p}}(\textbf{r},q_{\text{p}}) = \frac{d^2N}{ dq_{\text{p}} \ dV} = \Tilde{n}_{\text{p}}(\textbf{r}) \ q_{\text{p}}^{-\alpha_{\text{p}}} \ \theta(q_{\text{p}} - q_{\mathrm{min}}) 
\end{equation}
where $q_{\text{p}}$ is the dimensionless proton momentum defined as $q_{\text{p}}=\frac{P_{\text{p}}}{m_{\text{p}} c}$ and $m_{\text{p}}$ is the proton mass. $\theta(q_{\text{p}} - q_{\mathrm{min}})$ is the Heaviside step function, where $q_{\mathrm{min}}$ denotes the lower cut-off of the distribution function, and $\alpha_{\text{p}}$ is the CRp spectral index.

$\Tilde{n}_{\text{p}}(\textbf{r})$ denotes the normalization factor and exhibits number density dimensions $[\text{cm}^{-3}]$. In this work, $\Tilde{n}_{\text{p}}(\textbf{r})$ is the quantity that we consider as CRp density. In appendix \ref{conversion} we derive the relation between this normalization factor and the CRp flux that is the quantity quoted in direct CR measurements. 
\subsubsection{Pion-decay Induced Gamma-ray Emission}
The inelastic hadronic collision of CRps with the thermal ambient gas nucleons results in production of pions that decay to gamma-rays. In order to describe the hadronic proton-proton collision, there are two analytical models in the literature that assume isospin symmetry. The fireball model proposed by Fermi assumes thermal equilibrium of the pion cloud in the center of mass \citep{Fermi:1950jd}. In this model only CRps above a threshold $p_{\text{thr}} = 0.78 \ \text{GeV} \ c^{-1}$ can produce pions hadronically. In order to include more realistic effects such as the $\Delta_{\frac{3}{2}}$-isobaric model \citep{stecker1970cosmic} and the scaling model \citep{stephens1981production} near the pion-production threshold, we follow the analytical formalism given by \citet{2004A&A...426..777P}. This formalism is based on the approximate description of Dermer's model \citep{dermer1986binary, dermer1986secondary}.

Considering the power law density distribution of CRps and following the analytical formalism by \citet{2004A&A...426..777P} and \citet{ensslin2007}, the differential gamma-ray source function $s_{\gamma}(\textbf{r},E)$ is derived
\begin{equation}
    \begin{split}
        s_{\gamma}(\textbf{r},E) \ dE \ dV & =  \frac{2^4}{3 \alpha_{\gamma}}\ \tilde{n}_{\text{p}}(\textbf{r}) \ \frac{\sigma_{\text{p}\text{p}}} {m_{\text{p}} c} \ n_{\text{gas}}(\textbf{r}) \ \left( \frac{m_{\text{p}}}{2 m_{\pi^0}}\right)^{\alpha_{\gamma}} \\
    & \times \left[ \left( \frac{2E}{m_{\pi^0} c^2}\right)^{\delta_{\gamma}} + \left( \frac{2E}{m_{\pi^0} c^2}\right)^{-\delta_{\gamma}}\right]^{\frac{-\alpha_{\gamma}}{\delta_{\gamma}}} \ dE \ dV 
    \end{split}
\end{equation}
where $n_{\text{gas}}(\textbf{r})$ indicates the target nucleon density. The effective cross-section $\sigma_{\text{p}\text{p}}$ and the shape parameter $\delta_{\gamma}$ depend on the spectral index of the gamma-ray spectrum $\alpha_{\gamma}$ according to
\begin{equation}
    \sigma_{\text{p}\text{p}} \simeq 32 \ ( 0.96 + \text{e}^{4.4 \ - 2.4 \ \alpha_{\gamma} } ) \ \text{mbarn},
\end{equation}
and
\begin{equation}
    \delta_{\gamma} \simeq 0.14 \ \alpha_{\gamma}^{-1.6} + \ 0.44.
\end{equation}

According to Dermer's model and following the detailed discussion of \citet{2004A&A...426..777P}, the relation between the gamma-ray spectral index and CRp spectral index is given by $\alpha_{\gamma} = \alpha_{p}$. Furthermore, we assume $\alpha_{\text{p}} = 2.7$, consistent with local measurements of the cosmic-ray proton spectrum \citep{Aguilar2015} and with gamma-ray observations of molecular clouds probing the ambient CR population \citep{Aharonian2020,Peron21}.

In this model, the integrated gamma-ray source density $\lambda_{\gamma}(\textbf{r})$ is
\begin{equation}\label{integ gamma source density}
    \begin{split}
        \lambda_{\gamma}(\textbf{r}) & = \int_{E_1}^{E_2} \ dE \ s_{\gamma}(\textbf{r},E) \\
         & = \frac{4}{3 \ \alpha_{\gamma} \ \delta_{\gamma} } \  \tilde{n}_{\text{p}}(\textbf{r}) \ \frac{m_{\pi^0} \ c \  \sigma_{\text{p}\text{p}}} {m_{\text{p}}} \ n_{\text{gas}}(\textbf{r}) \ \left(\frac{m_{\text{p}}}{2 m_{\pi^0} }\right)^{\alpha_{\gamma}} \\
         & \times \left[\beta_{\text{x}}\left(\frac{\alpha_{\gamma} + 1}{2 \delta_{\gamma}} , \frac{\alpha_{\gamma} - 1}{2 \delta_{\gamma}} \right) \right]_{x_1}^{x_2} , \text{and}
    \end{split}
\end{equation}
\begin{equation*}
    x_i = \left[ 1 + \left( \frac{m_{\pi^0} c^2}{2 E_i}\right)^{ \delta_{\gamma}}\right]^{-1} \quad \text{for} \ \ i \in \{1,2\}.
\end{equation*}
Here $\beta_{\text{x}}(a,b)$ denotes the incomplete beta-function \citep{amiranoff2011abramowitz}.

We apply two standard conversion factors when relating the volumetric emissivity to the observed intensity. First, $\lambda_{\gamma}(\mathbf r)$ is defined as an omnidirectional production rate; the corresponding
gamma-ray intensity therefore includes the geometric factor $1/(4\pi)$ when integrating along the line of sight. Second, we account for the contribution of helium and heavier nuclei in both the cosmic-ray population and the interstellar medium by applying a nuclear enhancement factor. As mentioned earlier, we adopt the nuclear enhancement factor $\eta_{\gamma} \simeq 1.8$, consistent with estimates from \citet{Kafexhiu:2014cua} in the GeV energy range. With these considerations, we define the gamma-ray flux density by integrating $\lambda_{\gamma}(\textbf{r})$ over the line of sight (LOS)
\begin{equation}\label{gamma_flux}
    I_{\gamma}(x) = \frac{\eta_{\gamma}}{4\pi}\int d\ell \  \lambda_{\gamma}(\textbf{r}),
\end{equation}
where $x \in S^2$ denotes the position on the sky and $\textbf{r} = \ell \cdot x$.
\subsection{Cosmic Ray Proton Density Prior Model}\label{prior model}
\subsubsection{Bayesian Method}
We follow a probabilistic approach based on Bayesian inference to take into account the uncertainties. According to Bayesian inference, every possible configuration of the quantity of interest $s$ is considered and probability values are assigned to these configurations given the data $d$ via Bayes' theorem:
\begin{equation}
    \mathcal{P}(s|d) = \frac{\mathcal{P}(d|s) \ \mathcal{P}(s)}{\mathcal{P}(d)}.
\end{equation}
$\mathcal{P}(d|s)$ is the likelihood that indicates the probability of obtaining data $d$ for a given configuration of $s$. $\mathcal{P}(s)$ is the prior and accounts for our knowledge on $s$ a priori meaning pre-measurement. $\mathcal{P}(d) = \int ds \ \mathcal{P}(d|s) \ \mathcal{P}(s) $ is the evidence term that ensures the normalization of $\mathcal{P}(s|d)$ that is the posterior probability distribution.

\subsubsection{Prior Model}
Our quantity of interest is the CRp number density $\Tilde{n}_{p}(\textbf{r})$ that is the normalization parameter of the CRp distribution function described in eq.~\eqref{cr momentum dist}. By definition a density is a positive quantity and we assume that the CRp density is spatially smooth. A priori, we assume that the level of smoothness is spatially homogeneous and isotropic. We account for the positivity and smoothness of the CRp density by assuming that $\Tilde{n}_{p}(\textbf{r})$ is log-normally
distributed according to :
\begin{equation}
    \Tilde{n}_{p}(\textbf{r}) = \exp(s(\textbf{r}))
\end{equation}
where $s(\textbf{r})$ is normally distributed and is drawn from a Gaussian process with a homogeneous and isotropic kernel $K(\textbf{r})$.

Inspired by \citet{dust_edh}, we discretize our reconstructed 3D volume into HEALPix spheres that are logarithmically spaced in distance. We choose $N_{\text{side}}$ of 32 that correspond to 12288 plane of sky (POS) bins and we choose 196 logarithmically spaced distance bins that span from 0.05 kpc to 30 kpc from the Sun. The corresponding radial extension is chosen according Söding25 gas map's radial extension. The advantage of adapting a non linearly-spaced distance voxels is that we can investigate a larger volume while maintaining a high resolution at nearby distances. The discretized volume is defined heliocentrically.

Since our 3D modeled volume is discretized in inhomogeneously spaced voxels, we imprint the correlation structure using the Iterative Charted Refinement (ICR) method \citep{icr}. In ICR, the modeled volume is represented at multiple resolutions starting from a very coarse view and refining of the resolution is done iteratively until achieving the desired resolution. This technique has been successfully applied in similar Galactic tomography problems \citep{dust_edh,gas_25}.

The correlation structure that we consider for the CRp density is from the family of Matérn kernels \citep{2006gpml.book.....R}. The isotropic kernel is defined as the normalized correlation function obtained from a Matérn-like power spectrum,
\begin{equation}
K(r) = \frac{\xi(r)}{\xi(0)}, \qquad
\xi(r) = \frac{1}{2\pi^{2}}\int_{0}^{\infty} 
k^{2} P(k)\,\frac{\sin(kr)}{kr}\,dk,
\end{equation}
where we adopt the standard three-dimensional Fourier convention for isotropic fields.
with
\begin{equation}
P(k) = \frac{1}{
\bigl[1+(k\,\ell_{\mathrm{corr}})^{2}\bigr]^{\nu/2}}.
\end{equation}
where
$\ell_{\mathrm{corr}}$ is the correlation length that sets the characteristic scale of the kernel,
$\nu$ controls the logarithmic slope of the power spectrum,
and the normalization constant $\xi(0)$ ensures that $K(0) = 1$.

Both the correlation length $\ell_{\mathrm{corr}}$ and the spectral slope parameter $\nu$ are treated as scalar 
hyperparameters drawn from independent log-normal prior distributions. 
Specifically, each parameter $p \in \{\ell_{\mathrm{corr}}, \nu\}$ is modeled as
\begin{equation}
p \sim \mathrm{LogNormal}(\mu_p, \sigma_p),
\end{equation}
where $(\mu_p, \sigma_p)$ denote the mean and standard deviation of the corresponding log-normal prior in logarithmic space.

Furthermore, we include a multiplicative factor and an additive offset to model the CRp density. The multiplicative factor, drawn from a log-normal prior, captures deviations in the overall amplitude of the field, while the additive offset, drawn from a normal prior, allows flexibility in the baseline density. Our prior model for the CRp density is therefore
\begin{equation}
\tilde{n}_\mathrm{p}(\mathbf{r}) 
= \exp \bigl[ \mathrm{scl} \cdot s(\textbf{r}) + \mathrm{off} \bigr],
\end{equation}
where $\mathrm{scl}$ and $\mathrm{off}$ denote the multiplicative scale and additive offset, respectively. Our prior parameters are summarized in table\ref{PriorParams}. 
\begin{table*}
    \centering
    \setlength\extrarowheight{0.3ex}
    \caption{Parameters of the prior distributions. These parameters fully determine the prior for $\Tilde{n}_\mathrm{p}(\mathbf{r})$. The CRp reference density is $n_\mathrm{CR,Earth} = 6\times10^{-10}\,\mathrm{cm}^{-3}$.}
    \begin{tabular}{c|c|c|c|c|c}\toprule
        Parameter & Unit & Distribution & Mean & Standard deviation & Degrees of freedom \\\hline\hline
        $s$ & [1] & Normal & 0 & Matérn kernel & 2408448 \\
        $\mathrm{scl}$ & [1] & Log-normal & 1.5 & 3 & 1 \\
        $\nu$ & [1] & Log-normal & 6 & 2 & 1 \\
        $l_\mathrm{corr}$ & [kpc] & Log-normal & 6 & 3 & 1 \\
        $\mathrm{off}$ & [cm$^{-3}$] & Normal & $\ln(n_\mathrm{CR,Earth})$& 4 & 1 \\\hline
    \end{tabular}
    \label{PriorParams}
\end{table*}
\subsection{Likelihood and Noise Model}\label{likelihood}

To perform Bayesian inference we need to specify how the predicted gamma-ray emission relates to the observed data. The forward model described in Sect.~\ref{gamma-ray source model}
provides the expected gamma-ray flux density $I_{\gamma}(x)$ for a given CRp density field.
The likelihood then quantifies the probability of obtaining the observed gamma-ray map given this model prediction.

We adopt a Gaussian likelihood for the diffuse gamma-ray intensity map,
\begin{equation}
\mathcal{P}(d|s) =
\mathcal{N}\!\left(d - R(s),\, N\right),
\end{equation}
where $d$ denotes the observed gamma-ray intensity map, $R(s)$ represents the forward model that maps the CRp field $s$ to the predicted gamma-ray emission, and $N$ is the noise covariance matrix.

The noise level is estimated from the uncertainty map associated
with the diffuse gamma-ray component reconstruction of Scheel-Platz23. In our model we approximate the noise covariance as diagonal,
\begin{equation}
N_{ij} = \sigma_i^2 \, \delta_{ij},
\end{equation}
where $\sigma_i$ denotes the standard deviation of the diffuse
gamma-ray intensity in pixel $i$. This assumption neglects potential pixel-to-pixel correlations in the reconstructed gamma-ray map. Such correlations are expected to be present due to the component
separation procedure, but they are not explicitly modeled here and the noise is therefore treated as uncorrelated. As a consequence, the inferred uncertainties may be underestimated on large angular scales where correlated residual structures are present.

The use of a Gaussian likelihood is motivated by the nature of the input data. Rather than using raw photon counts, which would follow Poisson statistics, we employ the reconstructed diffuse gamma-ray
intensity map obtained by Scheel-Platz23. This map was itself inferred using \textit{Metric Gaussian Variational Inference} \citep[MGVI;][]{MGVI}, resulting in a field representation with approximately Gaussian uncertainties. Under this approximation, a Gaussian likelihood provides a consistent and computationally convenient description of the data uncertainties.
\subsection{Bayesian Inference Algorithm}\label{inference}
In previous sections, we have established our prior model describing the CRp density, as well as the model for deriving the expected $\gamma$-ray flux from the CRp interaction with the ISM. The integration of these components results in a forward model that maps the underlying physical field to the anticipated data. Our goal is to invert this forward model in order to infer the true physical state that produced the observed emission. In Bayesian terms, we aim to determine the posterior distribution $\mathcal{P}(s|d)$, which describes the most probable configuration of the CRp field given the data.

Direct sampling of the posterior distribution using methods such as Hamiltonian Monte Carlo \citep{1987PhLB..195..216D,2011hmcm.book..113N} is computationally infeasible due to the high dimensionality of the parameter space. Instead, we employ variational inference to approximate the posterior. Specifically, we use \textit{geometric Variational Inference} \citep[geoVI;][]{geovi}, a generalisation of MGVI \citep{MGVI}. In this approach, the posterior is represented by a multivariate normal distribution in a coordinate system induced by the Fisher information metric, which captures the local curvature of the posterior probability.

The algorithm proceeds iteratively. In each step, samples are drawn around an expansion point according to the local curvature of the posterior, approximated by the Fisher information. These samples provide a stochastic approximation to the true posterior. The expansion point is then updated by minimising the Kullback–Leibler divergence \citep{Kullback_1951} between the approximated and true posterior. This process is repeated until convergence, at which point the expansion point and local curvature become self-consistent. The resulting set of samples represents the posterior distribution and encapsulates the correlations between all model parameters.

The full model is implemented with the \texttt{NIFTy8} framework \citep{Selig2013,Steininger2019,Arras2019,Edenhofer2024}, which is based on \texttt{JAX} \citep{jax2018github}. This enables automatic differentiation and GPU-accelerated computations, facilitating efficient evaluation of the likelihood and its derivatives, as well as direct access to Fisher matrices without resorting to finite-difference methods.

Our reconstruction was performed on a single node of the Raven high-performance computing cluster at the Max Planck Computing and Data Facility (MPCDF), equipped with four NVIDIA A100 GPUs. The computation was carried out using a single GPU. During the inference, we used two posterior samples in the initial iterations and increased this number to twelve in later stages, with half of the samples serving as antithetical mirror counterparts.

\subsection{Validation on Synthetic Data}\label{validation}
\begin{figure*}[tbp]
    \centering
    \begin{subfigure}[b]{0.5\textwidth}
        \centering
        \includegraphics[width=\textwidth]{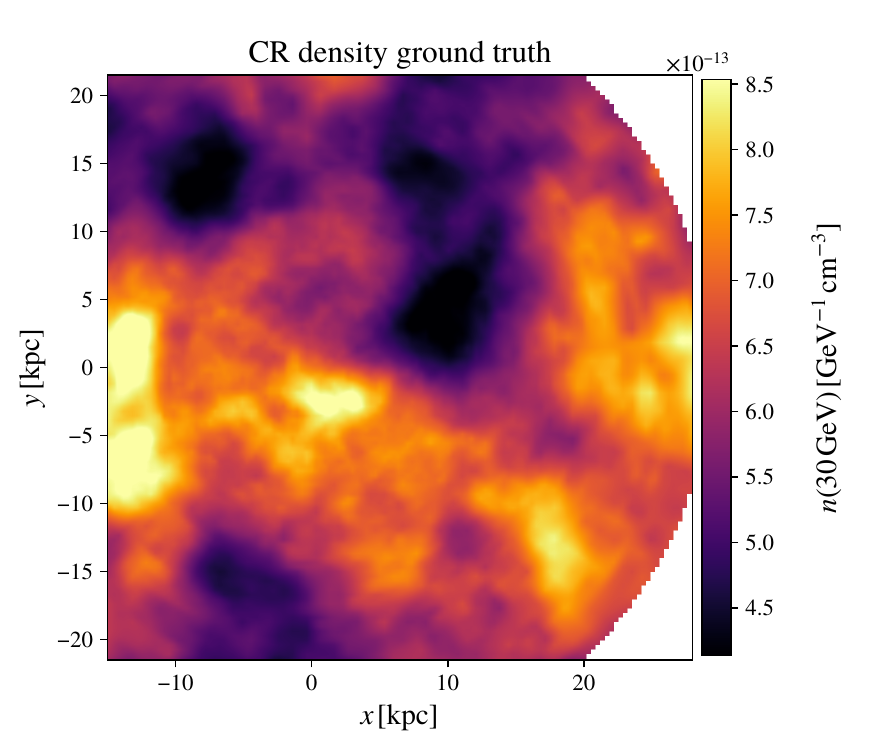}
        \label{fig:cr_truth}
    \end{subfigure}
    \hfill
    \begin{subfigure}[b]{0.48\textwidth}
        \centering
        \includegraphics[width=\textwidth]{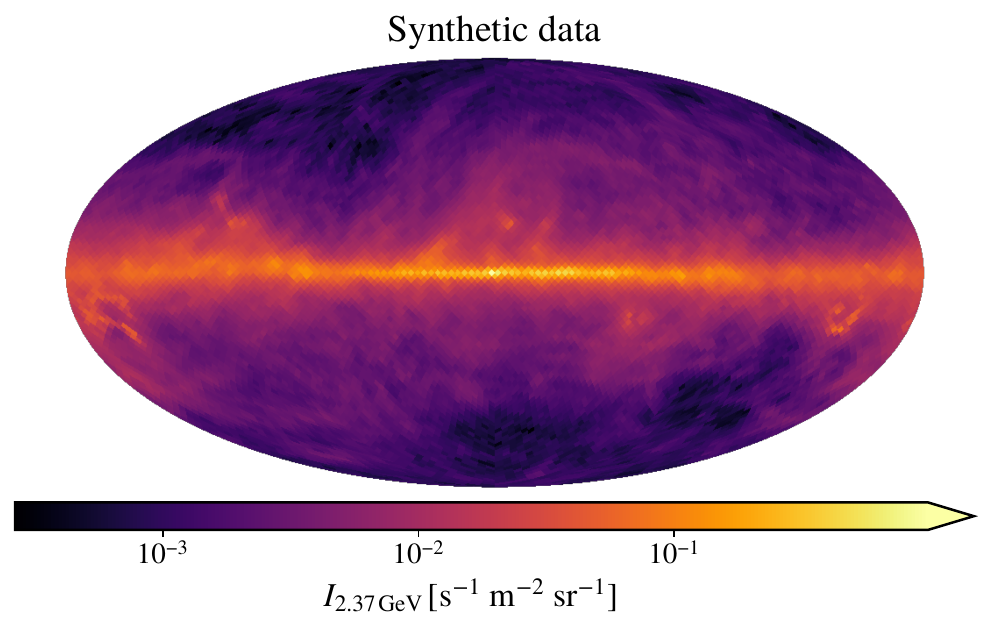}
        \label{fig:data}
    \end{subfigure}
    
    \vspace{0.3cm} 
    \begin{subfigure}[b]{0.5\textwidth}
        \centering
        \includegraphics[width=\textwidth]{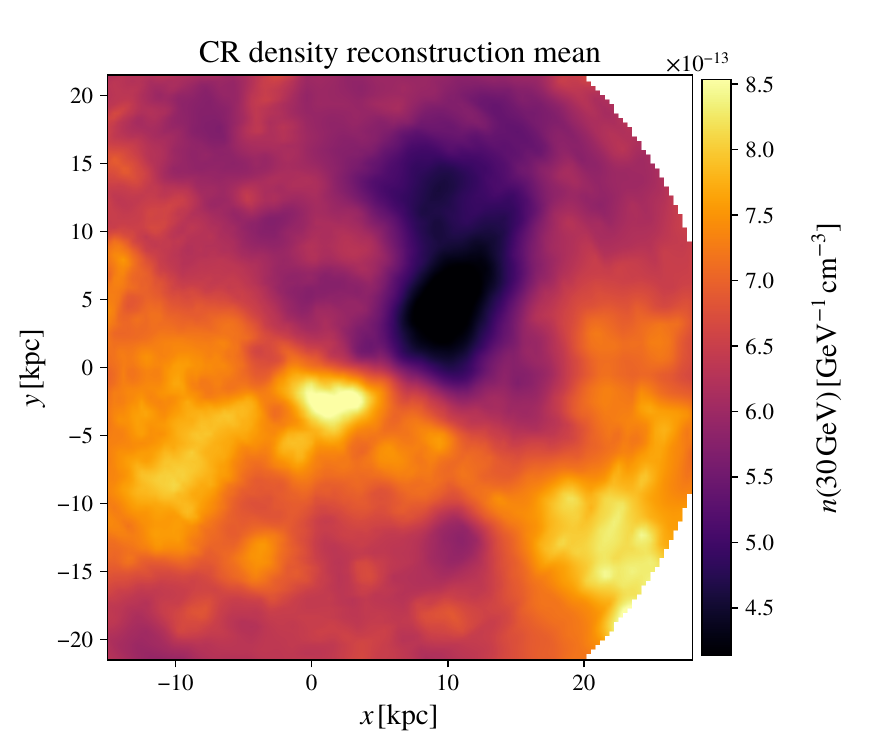}
        \label{fig:cr_rec}
    \end{subfigure}
    \hfill
    \begin{subfigure}[b]{0.48\textwidth}
        \centering
        \includegraphics[width=\textwidth]{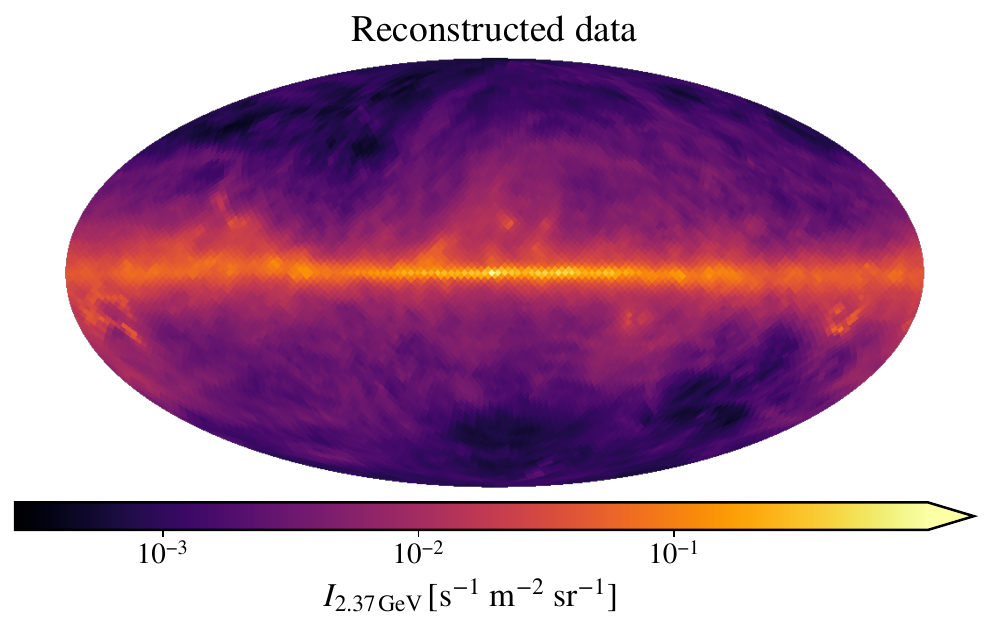}
        \label{fig:rec_data}
    \end{subfigure}

    \caption{Synthetic data validation. Top row: CR truth and synthetic gamma-ray data. Bottom row: reconstructed CR field and reconstructed gamma-ray data.}
    \label{cr_truth/rec}
\end{figure*}
\begin{figure*}[tbp]
    \centering
    \includegraphics[width=\linewidth]{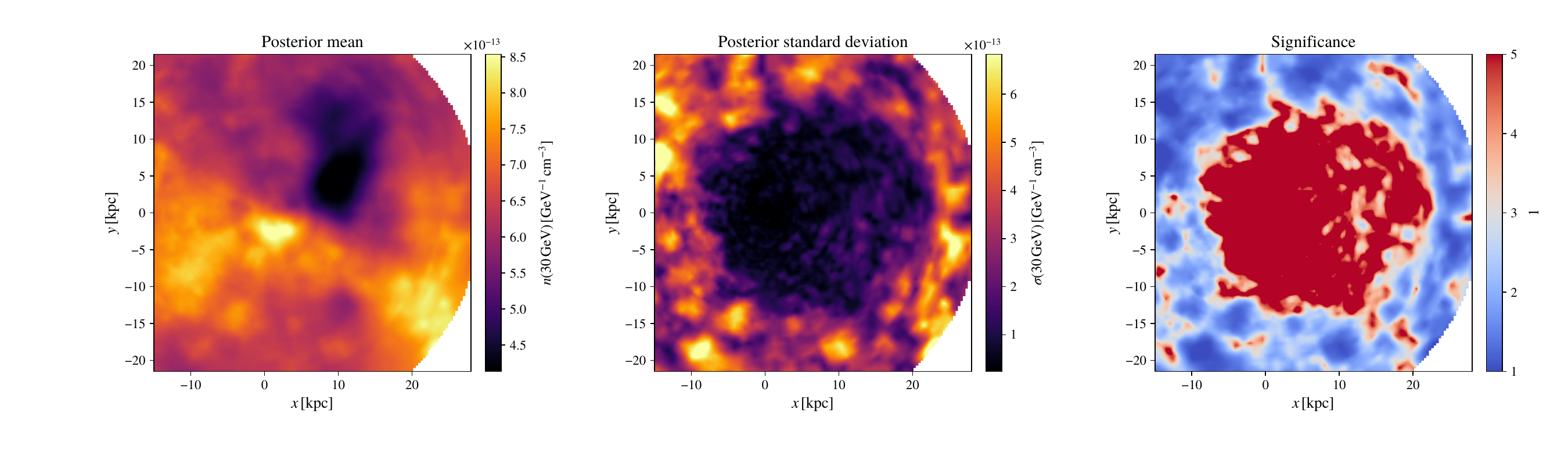}
    \caption{
       Synthetic data validation. Top-down view of the reconstructed cosmic-ray proton density in the 
        Galactic mid-plane ($z = 0$). The panels show (from left to right) the 
        posterior mean, posterior standard deviation, and the corresponding significance. 
        The significance is defined as the ratio of the posterior mean to the posterior 
        standard deviation. The mean and standard deviation maps are expressed in units 
        of $\tilde{n}_{\rm p}$ [cm$^{-3}$], while the significance map is dimensionless.
    }
    \label{fig:cr_xy_mean_std_snr}
\end{figure*}
To assess the performance and reliability of our inference pipeline, we first applied it to a synthetic data set for which the true CRp density field is known. This allows us to verify that the reconstruction accurately recovers the input field and to quantify potential biases or limitations inherent to our method. The synthetic data is generated by drawing a realization of the CRp density field from the prior model, which serves as the ground-truth distribution, and passing it through the forward model described in Sections \ref{gamma-ray source model} and \ref{prior model}, including the same gas map. Gaussian noise with a standard deviation matching the noise level of the gamma-ray map is added to the simulated data, ensuring that the synthetic dataset reflects the same observational conditions as the real data.

We then perform the full reconstruction procedure on this synthetic dataset using the same prior settings and algorithmic configuration as for the real analysis. This allows us to assess the algorithm’s ability to recover the true CRp field structure under realistic noise conditions. The comparison between the input (ground-truth) CRp field, the reconstructed field, and their corresponding gamma-ray maps provides a direct validation of both the inference method and the forward model consistency.

Figure \ref{cr_truth/rec} shows a comparison between the input (ground-truth) CRp field, the synthetic gamma-ray data (data), the reconstructed CRp field, and the corresponding reconstructed gamma-ray map. The ground-truth CRp map and the reconstructed field closely resemble each other in the overall morphology and large-scale structures, while the reconstructed gamma-ray map accurately reproduces the synthetic data. This confirms the consistency of the forward model and the effectiveness of the reconstruction pipeline in recovering the main features of the CRp distribution.

Figure \ref{fig:cr_xy_mean_std_snr} presents the posterior mean of the CRp density, its standard deviation, and the signal-to-noise ratio of the reconstruction. The posterior mean closely matches the ground-truth field, while the standard deviation and signal-to-noise maps indicate that the reconstruction is well constrained in regions with significant gas density. This demonstrates that our model is able to recover the main features of the CRp distribution under realistic noise levels. A caveat of the method is that it is primarily sensitive to regions where the gas density is non-uniform and sufficiently high; small-scale fluctuations in the CRp density or regions with very low gas density are less reliably reconstructed. Consequently, fine-scale structures in these regions may not be fully captured. Overall, the synthetic data validation confirms that the pipeline performs well and produces trustworthy results in the regions constrained by the available gas and gamma-ray data, as evidenced by the close agreement between the posterior mean and the ground-truth field and the high significance shown in Fig.~4.

\section{Results}

The results presented in this section are based on 12 posterior samples of the reconstructed normalization field $\tilde{n}_{\mathrm{p}}(\mathbf r)$ defined in Eq.~(\ref{cr momentum dist}). Throughout the following, we express this quantity in terms of the differential CRp number density evaluated at a reference kinetic energy of 30~GeV, $n(30\,\mathrm{GeV})$, which is obtained from $\tilde{n}_{\mathrm{p}}(\mathbf r)$ assuming a power-law CR spectrum with spectral index $2.7$. This choice follows common practice in the literature and facilitates comparison with previous studies \citep[e.g.,][]{Peron21,Gabici2022}, and is further motivated by the gamma-ray energy range considered here, for which the dominant hadronic emission is produced by CR protons in the tens-of-GeV regime. The corresponding conversion is described in Appendix~\ref{conversion}.\footnote{\url{https://doi.org/10.5281/zenodo.20328252}}

The reconstructed CRp maps are evaluated on a HEALPix $\times$ log($r$) grid. For $N_{\mathrm{side}}=32$, the effective angular resolution is $\sim1.8^{\circ}$ ($\sim110'$). The radial bin width varies with distance, ranging from $\sim8\,\mathrm{pc}$ in the innermost bins to $\sim0.8\,\mathrm{kpc}$ at the largest heliocentric distances, with a characteristic radial extent of $\sim130\,\mathrm{pc}$ at $r\simeq8\,\mathrm{kpc}$.
\begin{figure*}[tbp]
    \centering
    \begin{subfigure}[t]{0.49\hsize}
        \centering
        \includegraphics[width=\hsize]{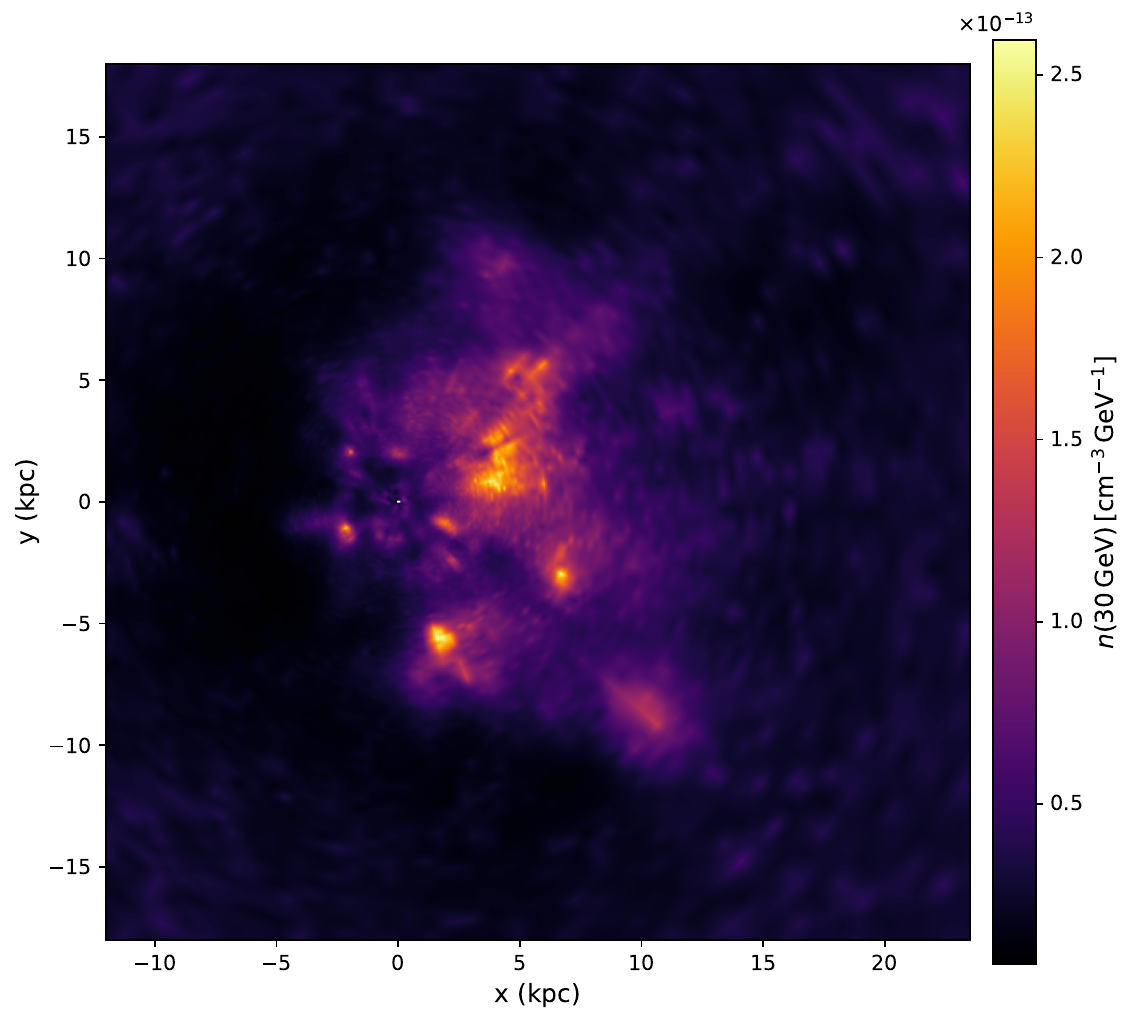}
        \caption{Mean cosmic-ray density, $n_{\rm CR,mean}(30\,\mathrm{GeV})$, at $z=0$.}
        \label{fig:cr_mean_z0}
    \end{subfigure}
    \hfill
    \begin{subfigure}[t]{0.49\hsize}
        \centering
        \includegraphics[width=\hsize]{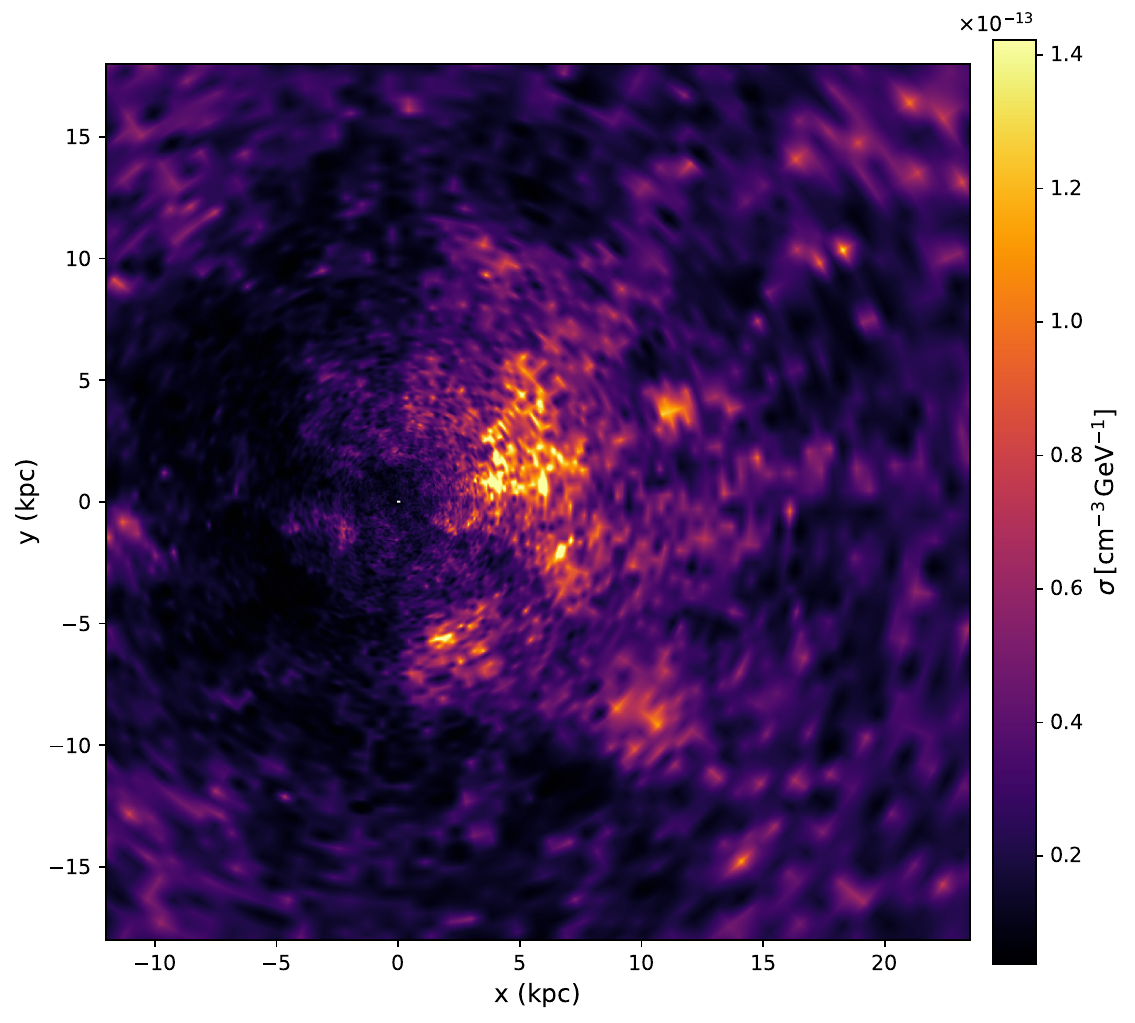}
        \caption{Reconstruction uncertainty, $\sigma_{\rm CR}(30\,\mathrm{GeV})$, at $z=0$.}
        \label{fig:cr_std_z0}
    \end{subfigure}

    \caption{
        Mid-plane ($z=0$) maps of the reconstructed cosmic-ray density.
        Left: posterior mean $n_{\rm CR,mean}(30\,\mathrm{GeV})$.
        Right: posterior standard deviation $\sigma_{\rm CR}(30\,\mathrm{GeV})$.
        The two panels use independent color scales to reflect their different dynamic ranges.
    }
    \label{fig:cr_mean_std_z0}
\end{figure*}
Figure~\ref{fig:cr_mean_std_z0} shows the posterior mean and posterior standard deviation of the reconstructed CRp density in a slice through the Galactic mid-plane ($z=0$). The maps are shown in heliocentric Cartesian coordinates, with the Sun located at $(x,y)=(0,0)$ and the Galactic center at approximately $(x,y)\simeq(8,0)$~kpc.

The posterior mean exhibits a moderate dynamical range across the Galactic plane, with CRp density variations remaining within approximately a factor of 10. The CRp density is distributed diffusely throughout the mid-plane rather than being dominated by a small number of isolated, high-contrast features. At the same time, moderate enhancements are visible toward the inner Galaxy, in particular within the Galactocentric range of approximately $4$--$6$~kpc and along lines of sight toward the Galactic center at larger heliocentric distances.

The corresponding posterior standard deviation shows a spatially varying uncertainty pattern, with elevated uncertainties in regions associated with these enhancements and at larger heliocentric distances. This indicates that, while the large-scale structure of the CRp distribution in the mid-plane is captured coherently, the level of constraint is not uniform across the reconstructed area.
\begin{figure*}[tbp]
    \centering
    \includegraphics[width=\textwidth]{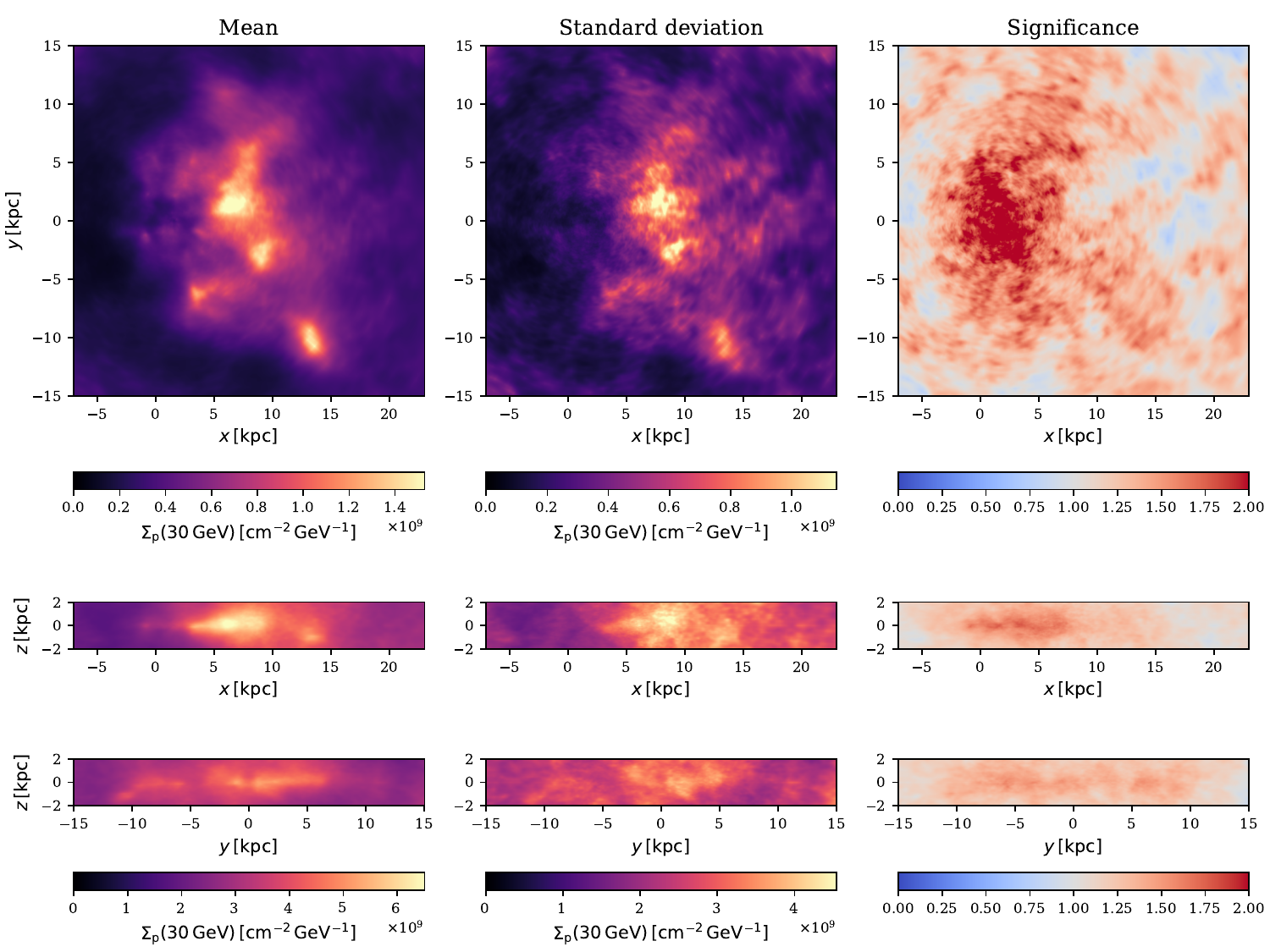}
    \caption{
        Axis-integrated views of the reconstructed CRp distribution along the Cartesian axes.
        From left to right, the panels show the posterior mean, posterior standard deviation,
        and the corresponding significance.
        The significance is defined as the ratio of the axis-integrated mean to the axis-integrated
        standard deviation.
        The top row shows quantities integrated along the $z$-axis and displayed in the
        $x$--$y$ plane, while the middle and bottom rows show quantities integrated along
        the $y$- and $x$-axes and displayed in the $x$--$z$ and $y$--$z$ planes, respectively.
    }
    \label{fig:cr_mean_std_snr_integrated}
\end{figure*}
\begin{figure*}[tbp]
    \centering
    \includegraphics[width=\textwidth]{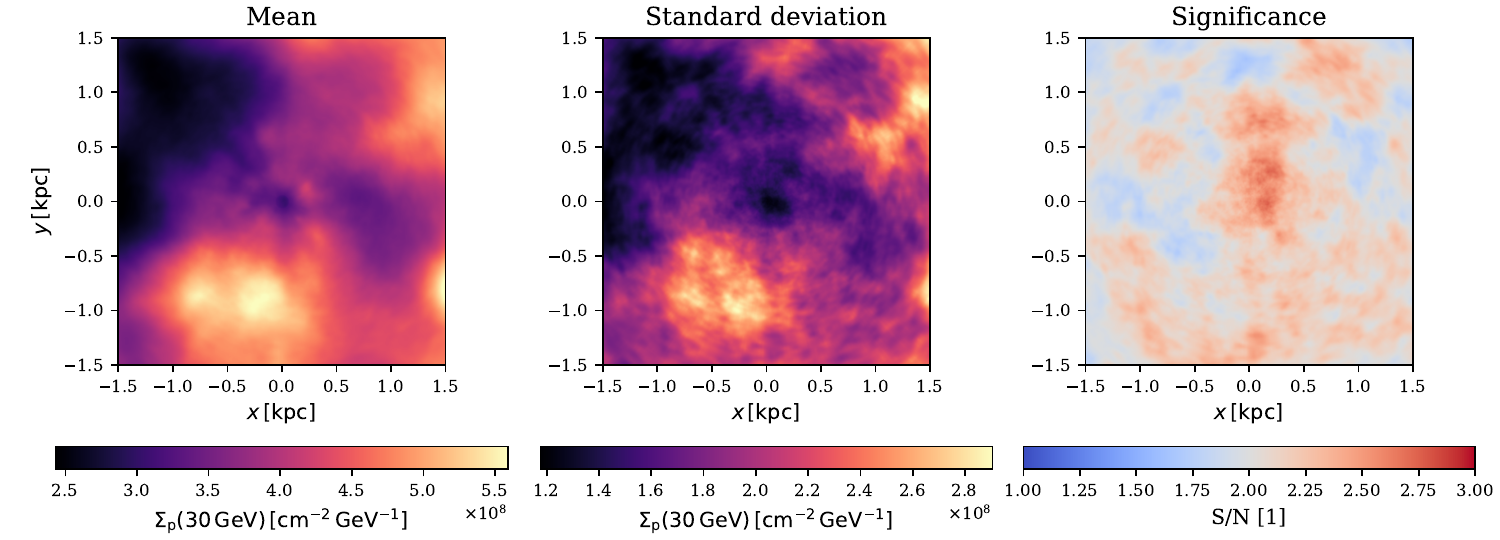}
    \caption{
        Zoomed-in $z$-integrated CRp density maps restricted to
        $x,y \in [-1.5,\,1.5]$~kpc in the vicinity of Earth.
        The panels show the posterior mean, posterior standard deviation,
        and the corresponding significance.
        The reconstructed local structure is primarily constrained by the
        3D dust map of \citet{dust_edh}, which extends to distances of
        approximately 1.25~kpc around the Sun.
    }
    \label{fig:cr_xy_zoom}
\end{figure*}

Figure~\ref{fig:cr_mean_std_snr_integrated} presents axis-integrated projections of the reconstructed CRp density along the Cartesian directions. For each projection, the posterior mean, posterior standard deviation, and the corresponding mean-to-standard-deviation ratio, hereafter referred to as the significance, are shown.

Compared to the mid-plane slice, the integrated projections appear smoother, as the integration accumulates contributions from extended regions along the respective axes. In the $x$–$y$ projection (top row), the enhancement in the Galactocentric range of approximately 4–6~kpc becomes more pronounced, reflecting its extended spatial extent.

The second and third rows show the $x$–$z$ and $y$–$z$ projections, respectively. In both cases, the CRp density is clearly confined to a disk-like structure around $z=0$, with a rapid decrease in amplitude toward larger vertical distances. The vertical extent of the reconstructed CRp distribution remains limited to a few kiloparsecs, and the disk structure is most prominent in the inner Galaxy. At larger heliocentric radii, the projected density becomes more diffuse and less structured.

The significance maps highlight regions where the integrated CRp density is well constrained relative to its posterior uncertainty. Higher significance values are predominantly found within the local and inner Galactic volume, where the gas distribution provides stronger constraints, while the outer regions exhibit lower significance. In the vertical projections, the highest significance follows the disk, indicating that the reconstruction is most reliable where the CRp density is spatially extended and supported by the underlying gas distribution.

Figure~\ref{fig:cr_xy_zoom} shows a zoomed-in view of the CRp density integrated along the $z$-axis and restricted to $x,y \in [-1.5,\,1.5]$~kpc. This region is chosen for visualization purposes and coincides with the spatial domain over which the 3D dust map by \citet{dust_edh} provides reliable constraints. The dust map is not used in the present reconstruction but motivates this choice of volume as a well-constrained local region for future studies.

In the posterior mean (left panel), the $z$-integrated CRp density exhibits smooth variations across the field with a limited dynamical range. The distribution remains spatially extended, without sharp discontinuities. A localized enhancement is visible in the lower-left quadrant, while the central region around $(x,y)\approx(0,0)$ appears comparatively less pronounced in this projection.

The posterior standard deviation (middle panel) follows the large-scale morphology of the mean, indicating that the uncertainty structure reflects the spatial distribution of the reconstructed CRp density rather than being spatially uniform.

The significance map (right panel), defined as the ratio of posterior mean to posterior standard deviation, shows systematically higher values than in the corresponding $z$-integrated projection over the full Galactic volume (Fig.~\ref{fig:cr_mean_std_snr_integrated}). Within this locally constrained region, the significance is predominantly above unity and reaches values of $\sim 2$--$3$ over extended areas. This indicates that the CRp density integrated along $z$ is more reliably constrained in the local volume than in the outer Galactic regions.
\begin{figure*}[tbp]
    \centering
    \begin{minipage}{0.49\textwidth}
        \centering
        \includegraphics[width=\linewidth]{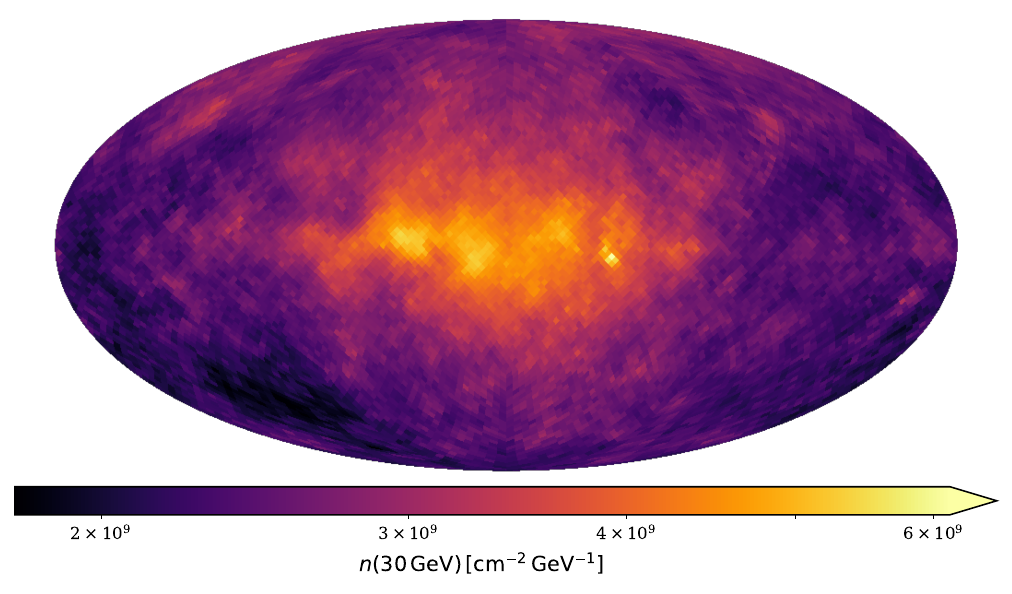}
        \caption*{(a) LOS-integrated CRp density, posterior mean}
    \end{minipage}
    \hfill
    \begin{minipage}{0.49\textwidth}
        \centering
        \includegraphics[width=\linewidth]{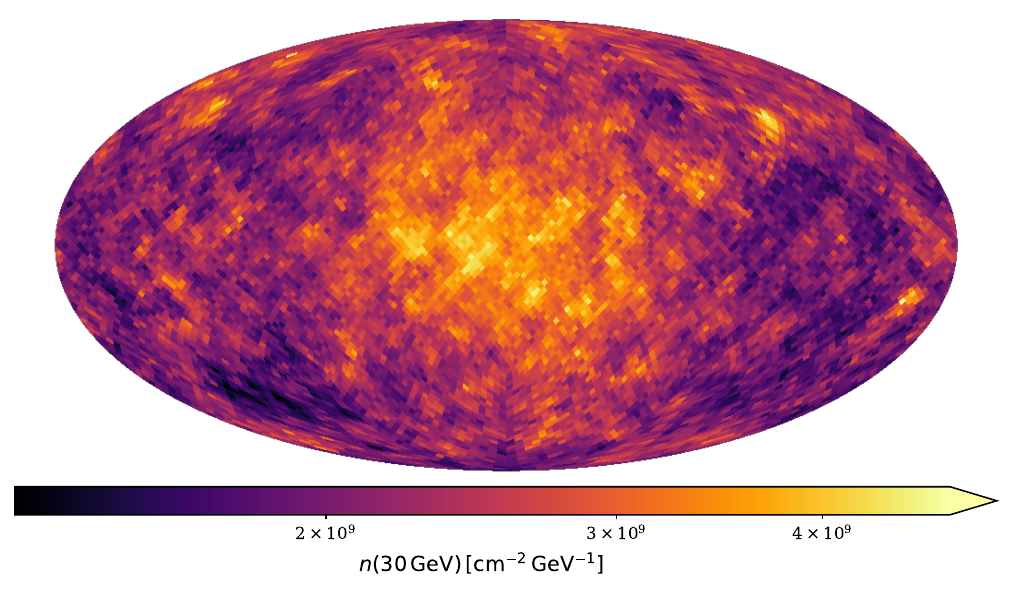}
        \caption*{(b) LOS-integrated CRp density, posterior standard deviation}
    \end{minipage}

    \caption{
        Plane-of-sky maps of the reconstructed CRp distribution obtained by
        integrating the three-dimensional CRp density along the line of sight.
        Panel (a) shows the posterior mean of the LOS-integrated
        $n(30\,\mathrm{GeV})$, while panel (b) shows the corresponding
        posterior standard deviation.
    }
    \label{fig:cr_los_mean_std}
\end{figure*}

Figure~\ref{fig:cr_los_mean_std} shows the plane-of-sky maps of the reconstructed CRp distribution obtained by integrating the three-dimensional CRp density along the line of sight. The left panel displays the LOS-integrated CRp density at a reference kinetic energy of 30~GeV corresponding to the posterior mean, while the right panel shows the associated posterior standard deviation.

Because the reconstruction yields the full three-dimensional CRp density field, this line-of-sight projection follows directly from the inferred volume distribution and provides the cumulative CRp content along each sky direction as seen from the Solar position.

The posterior mean map exhibits a pronounced enhancement along the Galactic plane, with the highest integrated CRp densities concentrated toward the inner Galaxy. This reflects the long sightlines through the Galactic disk in these directions. Away from the plane, the integrated CRp density decreases smoothly, consistent with the reduced path length through the disk and halo.

The posterior standard deviation shows a similar large-scale morphology, with enhanced uncertainties along the Galactic plane and toward the inner Galaxy. This indicates that the uncertainty is primarily driven by the spatial distribution of the reconstructed CRp density and the varying amount of constraining gas along different sightlines. At higher Galactic latitudes, both the integrated CRp density and its associated uncertainty decrease, reflecting the lower cumulative CRp content in these directions.

Together, these maps provide a self-consistent sky representation of the reconstructed CRp distribution and complement the Cartesian projections presented in Fig.~\ref{fig:cr_mean_std_snr_integrated}.

\section{Discussion}

The reconstruction presented in this work provides a three-dimensional view of the Galactic CRp density inferred from diffuse gamma-ray emission and a tomographic gas model. In this section, we discuss the physical implications of the recovered structures, the level of reliability across different regions of the Galaxy, and the limitations inherent to the adopted modeling assumptions.

\subsection{Large-scale CRp distribution}

Across all representations, including the mid-plane slice, the axis-integrated projections, and the plane-of-sky map, the CRp density exhibits a moderate dynamical range, with variations remaining within approximately a factor of 10 throughout the Galactic disk. This behavior is consistent with the picture of a quasi-uniform CR ``sea'' inferred from molecular cloud studies \citep{Peron21, Aharonian2020, Yang13} and with moderate enhancements toward the inner Galaxy indicated by diffuse gamma-ray analyses \citep{Yang16, Pothast2018}.

In particular, the Galactocentric range of approximately $4$--$6$~kpc shows a systematic enhancement relative to the local value. This feature is apparent both in the mid-plane slice and in the axis-integrated projections, indicating that it is spatially extended rather than arising from projection effects. A comparable enhancement in the molecular ring region has been reported in previous gamma-ray emissivity and molecular cloud studies (e.g., \citealt{Acero2016, Yang16, Pothast2018, Aharonian2020, Peron21}), which find elevated CR densities toward the inner Galaxy.

At the same time, the absence of extreme contrasts suggests that the large-scale CRp distribution is dominated by extended, diffusive transport rather than by a small number of localized, high-amplitude overdensities.

\subsection{Spatial reliability and reconstruction artefacts}

The significance maps presented in Fig.~\ref{fig:cr_mean_std_snr_integrated}
provide a quantitative indication of where the reconstruction is most reliable.
Regions with extended gas support and substantial integrated signal,
in particular the local and inner Galactic disk,
show systematically higher significance values,
whereas the outer Galaxy and regions with low gas density are less constrained.
\begin{figure*}[t]
    \centering
    \includegraphics[width=0.8\textwidth]{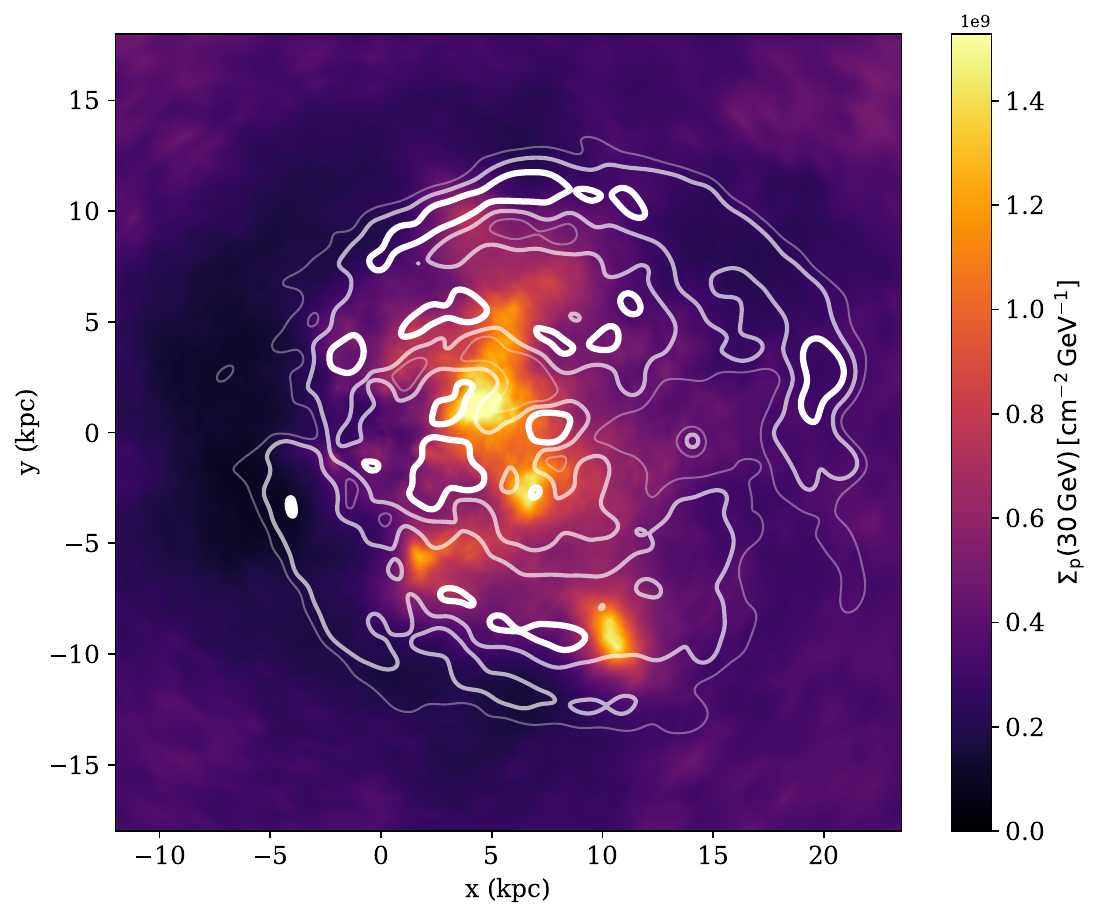}
    \caption{
        Vertically integrated CRp density at 30\,GeV in the $x$--$y$ plane
        with overlaid total gas surface-density contours.
        The color map shows $\int n_{\rm CR}(30\,\mathrm{GeV})\,\mathrm{d}z$,
        while the white contours trace the total gas surface density at
        $\Sigma_{\rm gas} = 2.47,\ 4.35,$ and $9.01\,M_\odot\,\mathrm{pc}^{-2}$,
        corresponding to the 60th, 75th, and 97th percentiles of the gas surface-density distribution, respectively.
        Regions of strong spatial correspondence indicate areas where the CR
        reconstruction is supported by substantial gas content.
    }
    \label{fig:cr_gas_overlay}
\end{figure*}

Figure~\ref{fig:cr_gas_overlay} illustrates this connection directly by showing the vertically integrated CRp density with overlaid gas morphology contours. In the inner Galactic disk, the reconstructed CR distribution follows the large-scale gas structure, indicating that the reconstruction is primarily driven by regions where the gamma-ray emission is well supported by the gas template. At larger Galactocentric radii, where the gas surface density decreases, the CR distribution becomes smoother and less structured, reflecting the reduced constraining power of the data.

A general feature of the reconstruction is the irregular spatial resolution imposed by the HEALPix $\times$ log($r$) grid. The radial bin width increases with heliocentric distance, leading to a gradual loss of small-scale structure at large radii. Consequently, fine-scale fluctuations in the outer Galaxy should not be over-interpreted.

In addition, some localized features, particularly in directions of complex gas morphology or near masked regions, may partially reflect residual modeling systematics. The masking of bright point-source regions and the Galactic center was introduced to mitigate known limitations of the diffuse emission modeling and the underlying gas map. Tests without these masks indicate that the reconstruction in these areas is not robust, and therefore these regions should be interpreted with caution.

While a broad correspondence between the integrated CRp density and the large-scale gas morphology is evident in Fig.~\ref{fig:cr_gas_overlay}, the alignment is not one-to-one. In particular, several regions with enhanced gas surface density do not show proportionally elevated CRp densities, and conversely, moderate CRp enhancements are visible in areas where the gas distribution is comparatively smooth. This indicates that the reconstructed CRp field is not merely tracing the gas template, but reflects spatial variations in the inferred CRp density.

Such deviations are expected because the gamma-ray emission depends on both the CR density and the gas density. Since the reconstruction combines the gamma-ray data with an assumed three-dimensional gas distribution, differences between the reconstructed CRp field and the gas morphology indicate that the inferred CRp distribution is not simply tracing the gas template.

\subsection{Hadronic gamma-ray source density versus gas column density}

\begin{figure*}[t]
    \centering
    \includegraphics[width=\textwidth]{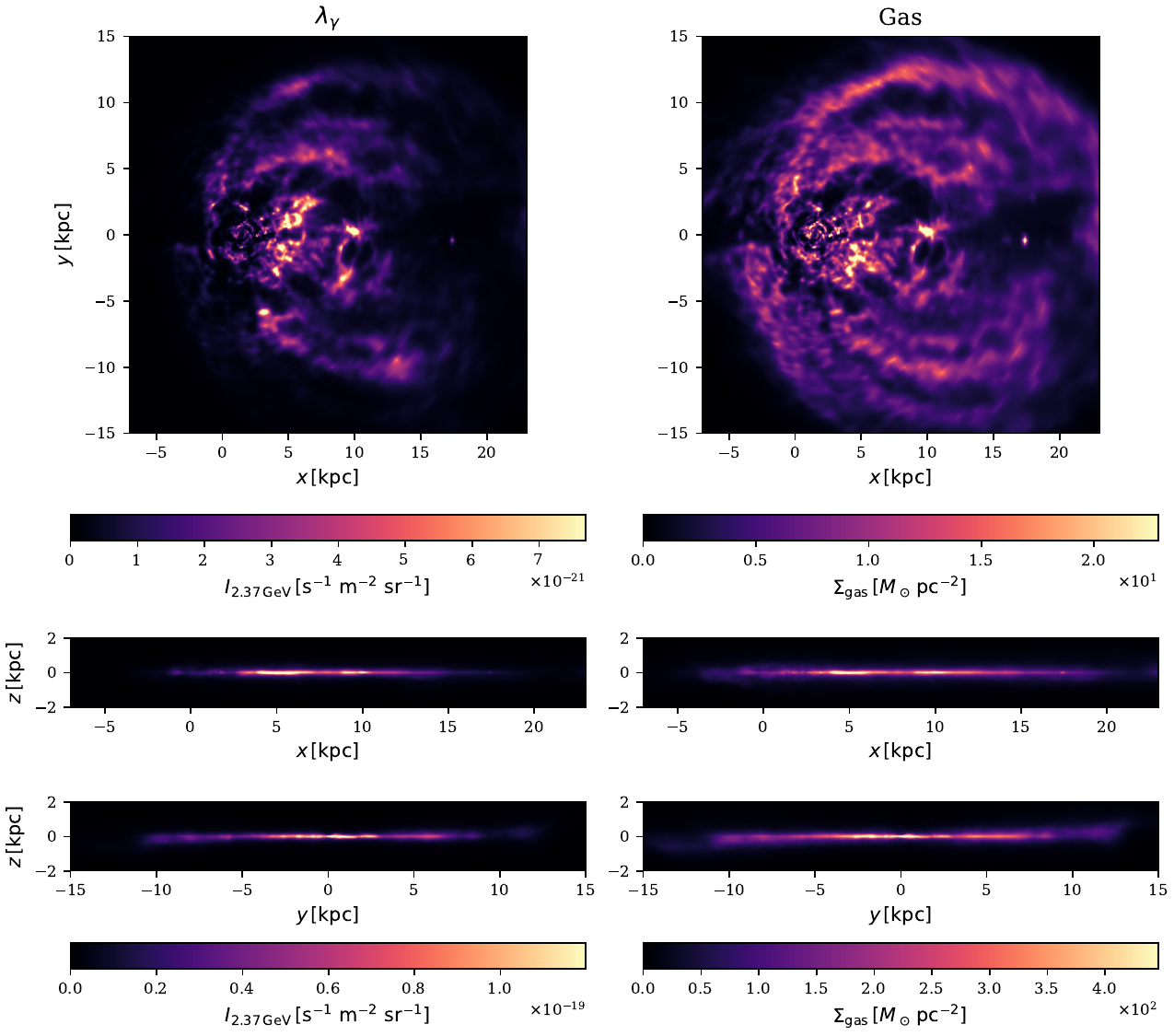}
    \caption{
        Comparison between projections of the hadronic gamma-ray source density $\lambda_{\gamma}$
        implied by the reconstruction and the corresponding gas column density.
        Left column: $\lambda_{\gamma}$ derived from the reconstructed CRp and gas density fields
        using the hadronic source model (Eq.~\ref{integ gamma source density}).
        Right column: total gas column density obtained from the 3D gas map.
        The top row shows projections along the $z$-axis in the $x$--$y$ plane, while the middle
        and bottom rows show projections along the $y$- and $x$-axes in the $x$--$z$ and $y$--$z$
        planes, respectively.
    }
    \label{fig:gamma_vs_gas}
\end{figure*}

Figure~\ref{fig:gamma_vs_gas} compares projections of the hadronic gamma-ray source density implied by our reconstruction with the corresponding projections of the gas distribution. The gamma-ray quantity shown here is derived from the reconstructed three-dimensional fields via the hadronic source model (Eq.~\ref{integ gamma source density}), i.e. $\lambda_{\gamma}(\mathbf r)$, which depends on the product of the CRp normalization field and the gas nucleon density. The gas panels show the corresponding projected gas column density obtained from the 3D gas map.

A broad morphological correspondence is visible between $\lambda_{\gamma}$ and the gas column density, reflecting the role of gas as the target material for hadronic interactions. At the same time, the correspondence is not one-to-one. Differences in contrast and spatial structure indicate that the morphology of the hadronic gamma-ray source density is not explained by the gas distribution alone, but is shaped by spatial variations in the reconstructed CRp field.

In the direction of the Galactic center, the comparison is affected by the rectangular mask applied to the gamma-ray data. As discussed in Sect.~\ref{data}, this region was excluded because of limitations in the underlying gas map and the complexity of the emission. Consequently, the reconstruction lacks direct constraints in this area, and the corresponding structures in $\lambda_{\gamma}$ should be interpreted with caution. The absence of structure in this region is therefore a deliberate modeling choice rather than a physical feature of the inferred CRp distribution.

\subsection{Comparison with previous measurements}
\begin{figure}[t]
    \centering
    \includegraphics[width=\columnwidth]{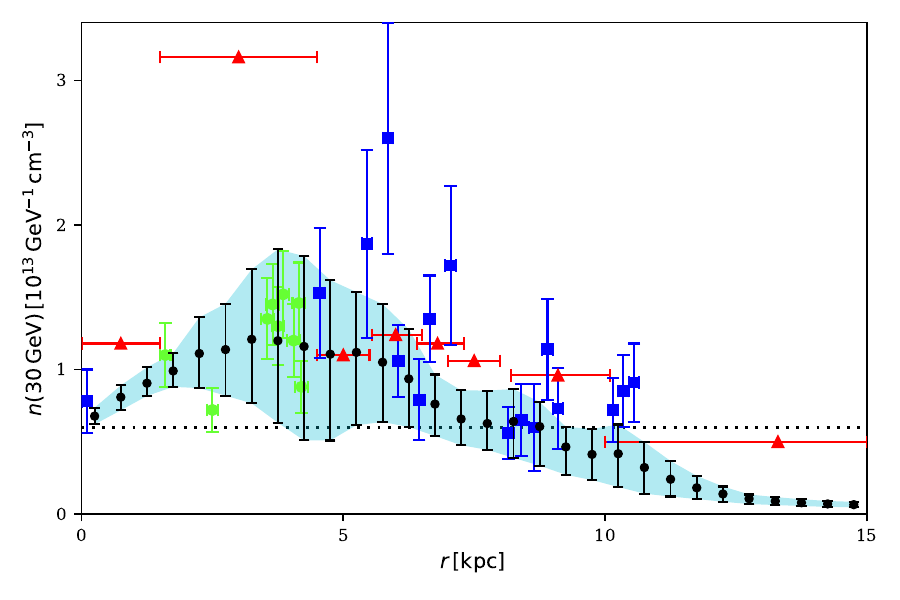}
    \caption{
        Normalisation of the CRp spectrum at 30~GeV as a function of Galactocentric distance as derived by gamma-ray observations.
        Red triangles, green circles, and blue squares show measurements from \citet{Acero2016}, \citet{Peron21}, and \citet{Aharonian2020}, respectively.
        The horizontal dotted line indicates the value measured locally by AMS--02 \citep{Aguilar2015}.
        Black points (with shaded band) show the posterior mean CRp density inferred in this work, with the band indicating the variations across positions at fixed Galactocentric radius.
        Figure adapted from \citet[see their Fig.~9, top panel]{Gabici2022}.
    }
    \label{fig:cr_radial_profile}
\end{figure}

Figure~\ref{fig:cr_radial_profile} compares the radial profile of the reconstructed CRp density at 30~GeV to previous determinations from gamma-ray emissivity and molecular cloud analyses. The literature measurements shown in the figure are taken from \citet{Acero2016}, \citet{Peron21}, and \citet{Aharonian2020}, and are displayed together with the local reference value inferred from AMS--02 \citep{Aguilar2015}. The figure layout follows the compilation presented by \citet[their Fig.~9, top panel]{Gabici2022}.

Our reconstructed profile (black points) lies close to the local reference level over a broad range of Galactocentric radii and shows a moderate enhancement in the inner Galaxy, in particular in the $R \simeq 4$--$6$~kpc range. Within uncertainties, the inferred trend is consistent with the spread of previous measurements in the same radial range, while remaining significantly below the most extreme values reported for individual regions. At larger Galactocentric distances, the reconstructed profile gradually decreases, in line with the decline reported in gamma-ray emissivity studies.

To assess the sensitivity of the reconstructed radial profile to the Galactic-center masking, we repeated the reconstruction using an alternative mask configuration. The resulting profile shows moderate quantitative differences in the inner Galaxy, as expected from the strong diffuse emission and increased modeling uncertainties toward the Galactic center. However, the broad enhancement in the Galactocentric range of approximately $4$--$6$~kpc remains present in both reconstructions. This indicates that the enhancement is not simply produced by edge effects associated with the adopted Galactic-center mask, but reflects a more extended feature of the reconstructed CRp distribution.

A key difference with respect to the quoted works is methodological: the present reconstruction does not impose a parametric radial profile or assume axisymmetry, but instead infers the full three-dimensional CRp normalisation field directly from the data, subject only to the smoothness assumptions encoded in the prior. The comparison in Fig.~\ref{fig:cr_radial_profile} therefore serves as an external consistency check that the large-scale radial behaviour recovered by the tomographic approach agrees with established gamma-ray-based constraints.

\subsection{Vertical structure and implications for CR transport}

The axis-integrated projections perpendicular to the Galactic plane (Fig.~\ref{fig:cr_mean_std_snr_integrated}) show that the reconstructed CRp distribution is confined to a disk-like structure around $z=0$, with a rapid decrease toward larger vertical distances. This behavior is consistent with a scenario in which CRs are continuously injected in the disk and subsequently diffuse into the halo.

\begin{figure}[t]
    \centering
    \includegraphics[width=\columnwidth]{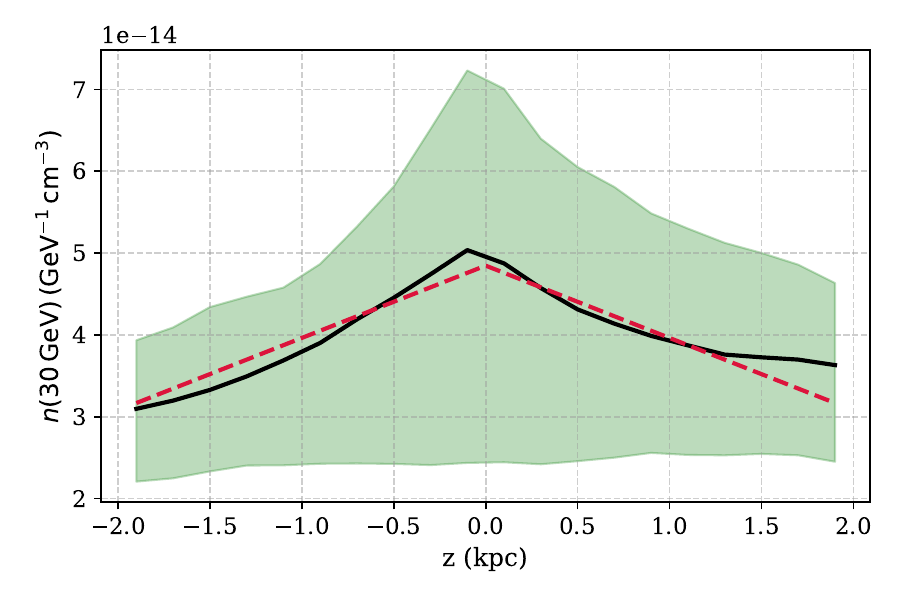}
    \caption{
    Vertical profile of the CRp density at 30~GeV in the Galactocentric radial range $8$--$9$~kpc. 
    The black line shows the posterior mean, and the shaded region indicates the $16$--$84\%$ credible interval. 
    The dashed red line corresponds to a linear fit of the form $a_1 - a_2 |z|$, used to characterize the vertical gradient and estimate an effective vertical extent of the reconstructed CR distribution.
    }
    \label{fig:cr_vertical_profile}
\end{figure}

To quantify this behavior, we extract the vertical profile of the CRp density in the radial range between 8 and 9~kpc, corresponding to a Galactocentric ring that passes through the solar neighborhood. This choice allows for a direct connection to the locally measured CR population and focuses on a region where the reconstruction is comparatively well constrained by both the gamma-ray data and the gas distribution. The resulting profile is shown in Fig.~\ref{fig:cr_vertical_profile}.

The profile exhibits a clear decrease with increasing $|z|$, consistent with the disk-dominated morphology inferred from the projected maps. To characterize this trend, we perform a simple symmetric fit of the form $a_1-a_2|z|$. Such an approximately linear decrease is qualitatively consistent with simple steady-state diffusion models with free escape boundaries, in which the CR density decreases linearly with distance from the Galactic plane. From this parameterization, we estimate an effective vertical extent of approximately $5.5~\mathrm{kpc}$ for the reconstructed CR distribution in the solar neighborhood. This quantity is intended only as a phenomenological characterization of the reconstructed profile and should not be interpreted as a direct measurement of the Galactic CR halo height.

The inferred vertical structure is influenced by the underlying gas distribution, the limited sensitivity of the gamma-ray data away from the disk, and the adopted spatial resolution, which becomes coarser at larger heliocentric distances. The effective vertical extent derived here should therefore be regarded as a constraint on the gamma-ray-traced CR distribution at 30~GeV rather than a determination of the full propagation halo size. Propagation studies based on diffuse gamma-ray observations commonly consider halo sizes of a few kiloparsecs, with values around $4~\mathrm{kpc}$ frequently adopted in the literature (e.g.~\citealp{Ackermann2012diffuse}). Within the uncertainties of the present reconstruction, the recovered profile is broadly consistent with such disk-confined CR transport scenarios. Nevertheless, the absence of strong high-$|z|$ enhancements suggests that the dominant CR population remains concentrated toward the Galactic disk within the sensitivity of the present analysis.

\subsection{Implications for the Galactic CR population}

Taken together, the reconstructed three-dimensional CRp distribution reveals a structured, yet moderately varying CR population across the Galactic disk. The density does not fluctuate by orders of magnitude, but neither is it strictly uniform. Instead, coherent variations on kiloparsec scales are visible in the mid-plane slice as well as in the axis-integrated projections.

The systematic enhancement in the Galactocentric range of approximately $4$--$6$~kpc, together with additional large-scale variations in the inner disk, indicates that the CRp distribution cannot be described by a purely constant ``sea''. At the same time, the absence of extreme, spatially confined overdensities suggests that transport smooths the CR distribution on sub-kiloparsec scales. The recovered morphology therefore supports a scenario in which a larger number of spatially located sources and large-scale propagation jointly shape the CR density profile.

Importantly, these structured features emerge without imposing a parametric radial model. The three-dimensional CRp normalization field is inferred directly from the gamma-ray data under minimal smoothness assumptions. The agreement of the recovered large-scale trends with previous emissivity-based studies therefore strengthens the interpretation
that the observed radial variations reflect genuine spatial modulation of the Galactic CR population rather than artifacts of simplified modeling.

A distinctive aspect of the present approach is that the CRp field is reconstructed in three spatial dimensions without enforcing axisymmetry. As a consequence, deviations from purely radial behaviour emerge naturally from the data rather than being imposed or suppressed by construction. The ability to derive both Cartesian projections and a self-consistent line-of-sight integrated map from the same underlying volume density highlights the added diagnostic power of the tomographic framework compared to traditional radially parameterized analyses.

\subsection{Model assumptions and limitations}

The present reconstruction relies on several simplifying assumptions that should be considered when interpreting the results.

First, the CRp spectral index is assumed to be spatially constant ($\alpha_{\mathrm p}=2.7$). Possible spatial variations of the CR spectrum are therefore not captured explicitly. If the spectral index varies across the Galaxy, part of the inferred normalization structure at 30~GeV could reflect spectral effects rather than purely amplitude variations.

Second, the analysis is restricted to a single gamma-ray energy range. While this choice reduces spectral complexity and mitigates contamination from additional emission components, it does not allow us to probe energy-dependent spatial variations of the CR population.

Third, only hadronic gamma-ray emission is modeled explicitly. Residual contributions from other processes, if not fully removed in the diffuse component separation, may introduce localized systematics in the inferred CRp normalization.

Finally, the reconstruction depends on the adopted three-dimensional gas model. Systematic uncertainties or structural limitations in the gas tomography propagate directly into the CRp field. Although masking was applied in regions where the gas constraints are known to be unreliable, improvements in future gas maps will directly enhance the robustness of tomographic CR reconstructions.

\section{Conclusion and Outlook}

We have presented a fully three-dimensional, data-driven tomographic reconstruction of the cosmic-ray proton (CRp) density in the Milky Way, inferred from diffuse gamma-ray emission and a Bayesian 3D gas model. The reconstruction is represented in terms of the CRp normalization field and expressed as a spatially resolved differential CRp number density at a reference kinetic energy of 30~GeV, assuming a power-law CR spectrum with spectral index 2.7, without imposing a predefined radial profile or axisymmetric structure.

The recovered CRp distribution exhibits coherent kiloparsec-scale variations across the Galactic disk, including a moderate enhancement in the Galactocentric range of approximately $4$--$6$~kpc. At the same time, the overall dynamical range remains moderate, indicating a structured but not strongly fragmented CR population. The vertical projections indicate confinement of the dominant CR component to the Galactic disk within the sensitivity and resolution limits of the present reconstruction.

Comparison with previous gamma-ray emissivity measurements shows good agreement in the large-scale radial trend, providing an independent validation of earlier results within a fully three-dimensional, non-parametric framework. The tomographic approach further reveals regions where spatial variations in the CR population cannot be explained by the gas distribution alone, indicating genuine structure in the CR density field. Importantly, such variations are also observed in nearby regions where the gas distribution is comparatively well constrained, suggesting that these features are unlikely to arise solely from uncertainties in the gas model.

The present analysis is subject to several simplifying assumptions, including a spatially constant CR spectral index, restriction to a single gamma-ray energy range, and reduced sensitivity in regions where the gas distribution is smooth or of low density. Future improvements in three-dimensional gas maps and extensions to multi-energy gamma-ray analyses will enable a more complete reconstruction of both the CR density and spectral structure across the Galaxy.

More broadly, this work demonstrates that three-dimensional cosmic-ray tomography based on diffuse gamma-ray emission is not only feasible, but capable of recovering physically meaningful spatial structure beyond simple radial or axisymmetric parameterizations. The framework developed here provides a foundation for future studies that aim to combine gamma-ray observations, gas tomography, and CR transport models within a unified three-dimensional description of the Galactic CR population, representing an important step toward data-driven constraints on Galactic CR transport.
\begin{acknowledgements}
Some of the results presented in this paper were obtained using the healpy package and the HEALPix software. Several figures were produced using Matplotlib (Hunter 2007). This work was supported by the Deutsche Forschungsgemeinschaft (DFG, German Research Foundation) under project number 495252601. Part of this research was funded by the Austrian Science Fund (FWF) under grant I 5925-N. H.Z. gratefully acknowledges fruitful discussions with Vincent Eberle.
\end{acknowledgements}
\bibliographystyle{aa}
\bibliography{bib}
\appendix
\section{Reconstruction using only the LMC/SMC mask}\label{appendix_gc_mask}

In the main analysis (Sect.~\ref{data}), we apply a composite mask that excludes
the LMC and SMC, regions associated with bright pulsars and strong template
modification, as well as a rectangular region around the Galactic centre.
In this appendix we repeat the reconstruction using a minimal mask in which
only the LMC and SMC are excluded. All other settings, including the gas map,
energy range, prior configuration, and inference procedure, are kept identical
to the main analysis.

\begin{figure*}[tbp]
    \centering
    
    \begin{subfigure}[t]{0.49\textwidth}
        \centering
        \includegraphics[width=\linewidth]{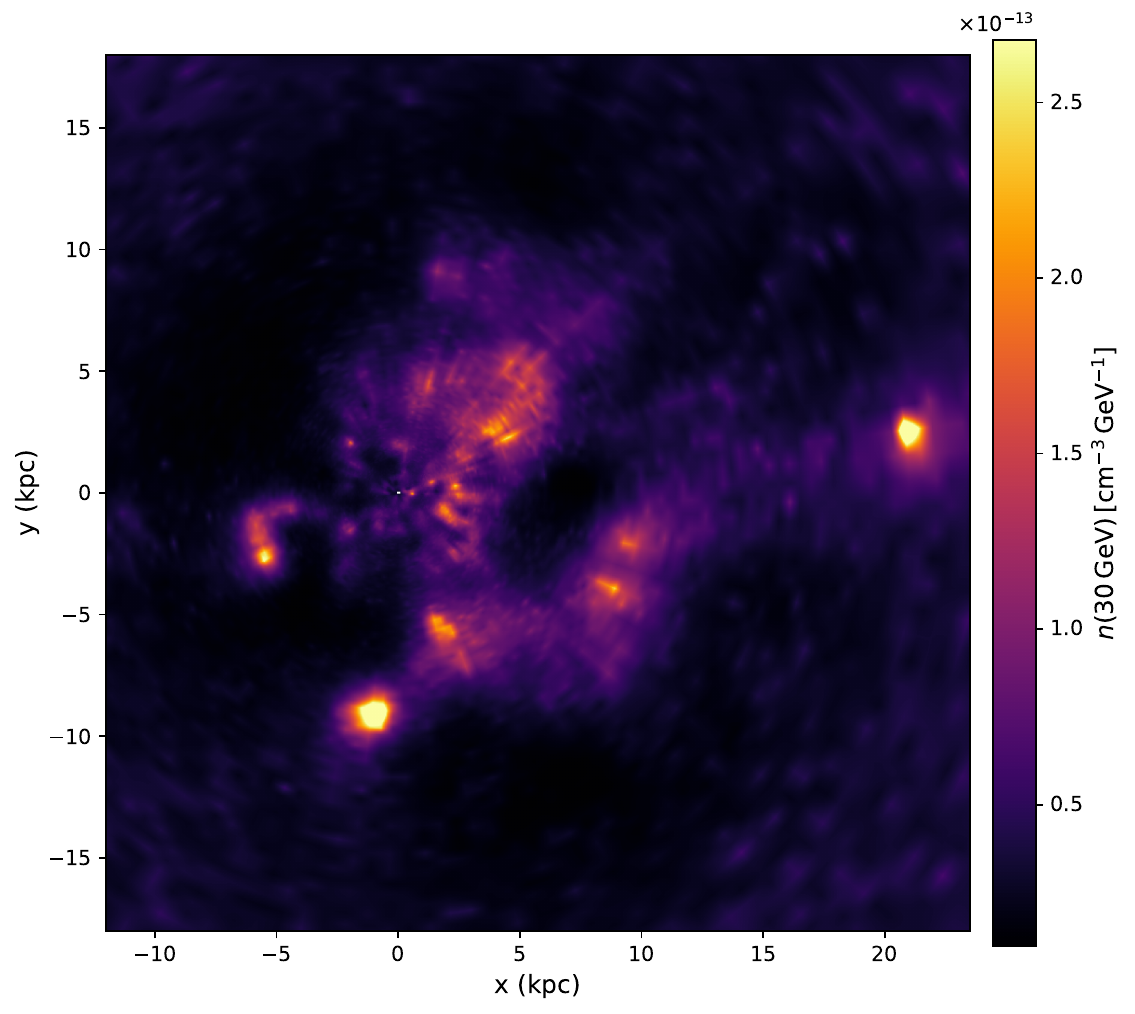}
        \caption{Posterior mean at $z=0$ (only LMC/SMC masked).}
        \label{fig:cr_mean_z0_minmask}
    \end{subfigure}
    \hfill
    \begin{subfigure}[t]{0.49\textwidth}
        \centering
        \includegraphics[width=\linewidth]{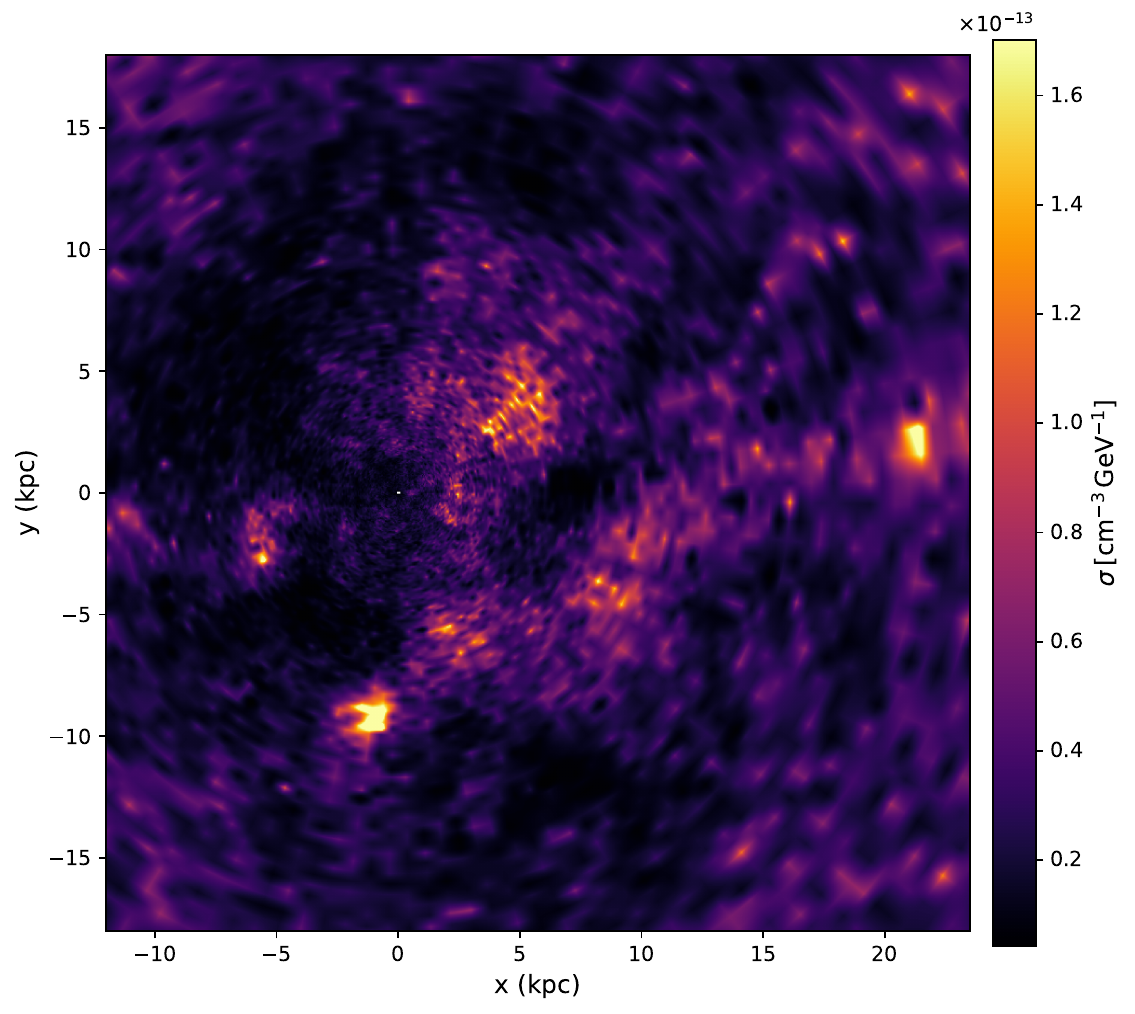}
        \caption{Posterior standard deviation at $z=0$ (only LMC/SMC masked).}
        \label{fig:cr_std_z0_minmask}
    \end{subfigure}
    
    \vspace{0.8cm}
    
    \includegraphics[width=\textwidth]{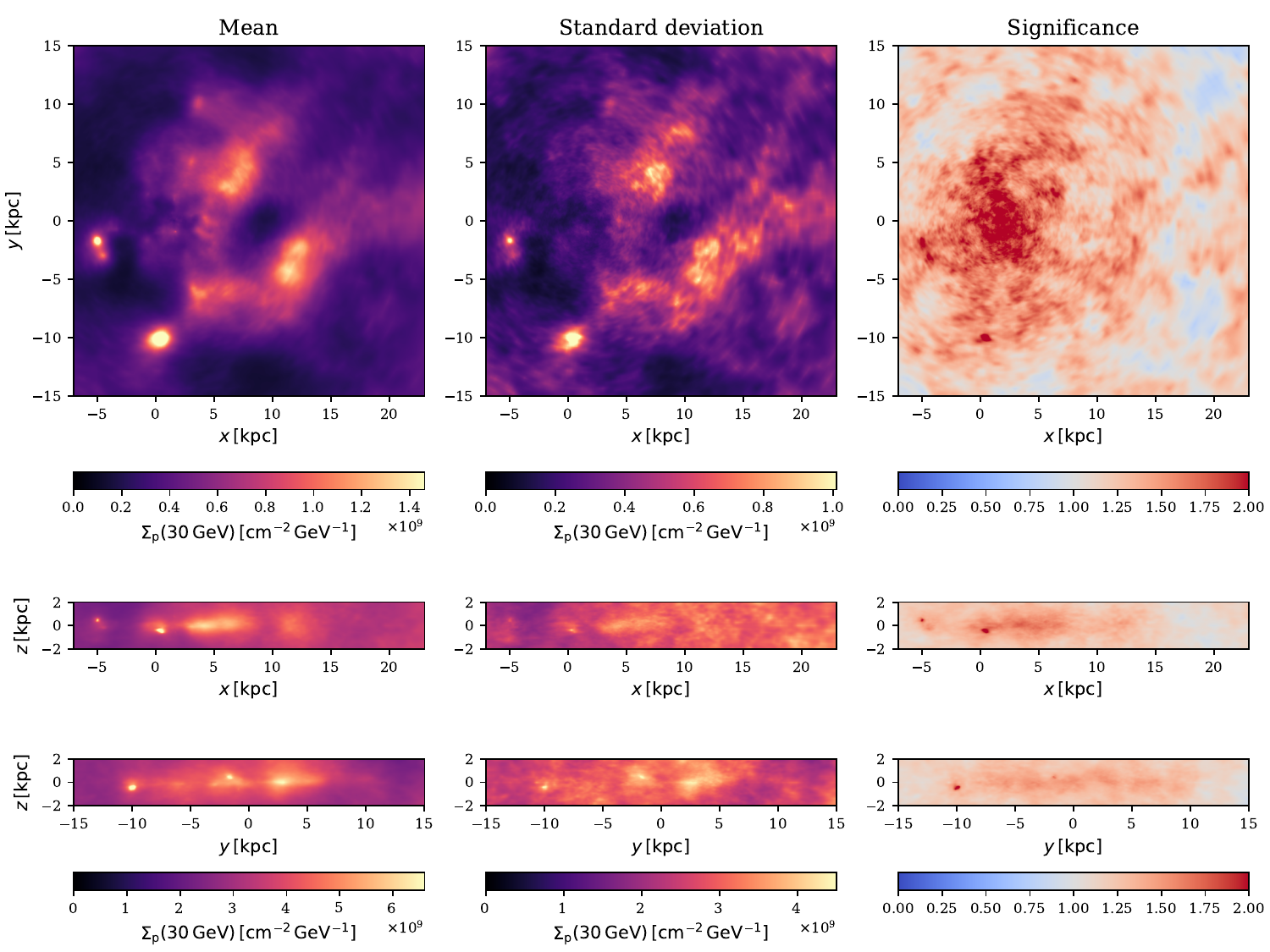}
    
    \caption{
        Reconstruction using only the LMC and SMC mask.
        \textbf{Top:} Mid-plane ($z=0$) posterior mean and standard deviation.
        \textbf{Bottom:} Axis-integrated projections (mean, standard deviation, and significance).
        Compared to the main analysis, the vertical projections exhibit irregular structure and enhanced uncertainties toward the Galactic-centre direction.
    }
    \label{fig:cr_minmask_all}
\end{figure*}
The comparison demonstrates that including the Galactic-centre region
and pulsar-dominated sky areas materially affects the stability of the
three-dimensional reconstruction. In particular, the $x$--$z$ and $y$--$z$
projections show less regular vertical structure and elevated posterior
variance in the inner Galaxy. These differences indicate sensitivity to
directions where the diffuse emission separation and/or the gas tomography
provide weaker constraints. For this reason, the additional masking adopted
in the main analysis improves the robustness of the inferred large-scale
CRp distribution.

\section{{Conversion of CRp flux into the CRp normalization factor $\Tilde{n}_{\text{p}}(\textbf{r})$}}\label{conversion}
Here we derive the relation between the normalization factor $\tilde{n}_{\mathrm{p}}$ following \citet{ensslin2007} and the CR flux, which is typically quoted in direct CR observations. Independently of the specific definition, the integrated distribution yields the CR number density. We therefore have
\begin{eqnarray}
    \tilde{n}_{\mathrm{p}}\left(\frac{p}{m_{\mathrm{p}}c}\right)^{-\alpha}\frac{{\rm d}p}{m_{\mathrm{p}}c} = \frac{4\pi j(E)}{v} {\rm d}E,
\end{eqnarray}
where $j(E)$ denotes the differential CR flux and $v$ the particle velocity. This leads to
\begin{eqnarray}
    \tilde{n}_{\mathrm{p}} = 4\pi j(E) \left(\frac{p}{m_{\mathrm{p}}c}\right)^{\alpha} m_{\mathrm{p}}c.
\end{eqnarray}

In the local ISM, we can estimate $\tilde{n}_{\mathrm{p}} \simeq 6 \times 10^{-10}\,{\rm cm}^{-3}$. The differential number density of CRs follows from
\begin{eqnarray}
    \tilde{n}_{\mathrm{p}}\left(\frac{p}{m_{\mathrm{p}}c}\right)^{-\alpha}\frac{{\rm d}p}{m_{\mathrm{p}}c} = n(E)\,{\rm d}E,
\end{eqnarray}
which yields
\begin{eqnarray}
    n(E) = \tilde{n}_{\mathrm{p}}\left(\frac{p}{m_{\mathrm{p}}c}\right)^{-\alpha}\frac{1}{m_{\mathrm{p}}c^2 \beta}.
\end{eqnarray}

For consistency with the analysis in the main text, we evaluate the differential CR number density at a reference energy of $E=30$~GeV. This choice is motivated by the energy range of the gamma-ray data considered in this work, for which the dominant hadronic emission probes CR protons in the tens-of-GeV regime, and follows common conventions adopted in molecular cloud studies.

For $E=30$~GeV and assuming $\alpha \simeq 2.7$, we obtain the approximate relation
\begin{eqnarray}
    n(30\,{\rm GeV}) \simeq 8.5\times 10^{-5} \left(\frac{\tilde{n}_{\mathrm{p}}}{1\,{\rm cm}^{-3}}\right)\, {\rm GeV}^{-1}\,{\rm cm}^{-3}.
\end{eqnarray}

Using $\tilde{n}_{\mathrm{p}} \simeq 6\times10^{-10}\,{\rm cm}^{-3}$ for the local ISM, this corresponds to
\begin{eqnarray}
    n(30\,{\rm GeV}) \simeq 0.5\times 10^{-13}\,{\rm GeV}^{-1}\,{\rm cm}^{-3},
\end{eqnarray}
which is consistent with the results shown, for example, in Fig.~9 of \citet{Gabici2022}. 
\end{document}